\documentclass[twocolumn,amsthm]{autart}

\usepackage{amsmath,amssymb}
\usepackage{booktabs,array}
\usepackage{tikz}
\usepackage{subcaption}
\usepackage{graphicx}
\usepackage{xcolor}
\makeatletter\def\ps@copyright{\let\@mkboth\@gobbletwo\def\@oddhead{}\def\@evenhead{}\def\@oddfoot{}\def\@evenfoot{}}\makeatother
\makeatletter
\renewcommand\paragraph{\@startsection{paragraph}{4}{\z@}{0.0001pt}{-1em}{\normalsize\bfseries}}
\makeatother
\newcommand{\nomgroup}[2]{\par\vskip3pt\hrule height #1\nobreak\vskip4pt\nobreak
 \noindent\emph{#2}\par\nobreak\vskip3pt\nobreak\hrule height .3pt\nobreak\vskip4pt\nobreak}
\newcommand{\nomitem}[2]{\par\noindent\hangindent=0.24\columnwidth\hangafter=1
 \makebox[0.24\columnwidth][l]{$#1$}\ignorespaces#2\par\vskip2pt}

\usepackage[authoryear,round]{natbib}
\usetikzlibrary{calc,arrows.meta}
\makeatletter\@ifundefined{c@part}{\newcounter{part}}{}\makeatother
\usepackage[colorlinks=true,linkcolor=blue,citecolor=blue,urlcolor=blue]{hyperref}

\makeatletter\g@addto@macro\endfrontmatter{\endNoHyper}\makeatother

\theoremstyle{plain}
\newtheorem{theorem}{Theorem}

\newtheorem{proposition}{Proposition}
\newtheorem{corollary}{Corollary}
\newtheorem{definition}{Definition}
\newtheorem{assumption}{Assumption}
\theoremstyle{remark}
\newtheorem{remark}{Remark}

\protected\def\Krstic{Krsti\'c}
\newcommand{\atantwo}{\operatorname{atan2}}

\tikzset{>={Latex[length=2.2mm]},
  ang/.style={gray!70,line width=0.5pt},
  ref/.style={gray!70,dashed,line width=0.5pt},
  leg/.style={line width=1.1pt,->},
  hull/.pic={\draw[fill=#1] (-0.90,-0.28)--(0.55,-0.28)--(0.95,0)--(0.55,0.28)--(-0.90,0.28)--cycle;}}

\allowdisplaybreaks
\begin{document}

\begin{frontmatter}

\title{Control of Underactuated Surface Vessels---Part I:
%\\ Underway Docking
Stabilization 
%of Underactuated Surface Vessels\\ 
in Polar Coordinates
\thanksref{footnoteinfo}}

\thanks[footnoteinfo]{This paper was not presented at any IFAC meeting. Corresponding author M.~Krsti\'c.}

\author[NTNU]{Kristin Y. Pettersen}\ead{kristin.y.pettersen@ntnu.no},
\author[UCSD]{Miroslav \Krstic}\ead{krstic@ucsd.edu}

\address[NTNU]{Department of Engineering Cybernetics, Norwegian University of Science and Technology (NTNU), Trondheim, Norway}
\address[UCSD]{Department of Mechanical and Aerospace Engineering, University of California San Diego, La Jolla, CA 92093, USA}


\begin{abstract}
A surface vessel with surge force and yaw moment as inputs and no sway
actuation cannot be stabilized at a fixed pose by continuous time-invariant
feedback. We introduce a \emph{zero-sideslip point}---a point whose position
relative to the hull depends on the surge and sway speeds and whose velocity
points along the vessel's heading at every instant---and show that it moves as
a unicycle. This reduction allows the time-invariant, polar-coordinate
steering laws of Todorovski et al.\ (2025) to be applied to the
underactuated vessel without modification. With two steering laws---GloBa and
BoLSA---we construct strict control Lyapunov functions and prove global
asymptotic stability, first for a four-state model with surge speed and yaw
rate as virtual inputs, and then, by one step of backstepping, for the full
six-state model with surge force and yaw moment. The analysis extends to
hulls with fore-aft asymmetric sway--yaw damping, excluded in prior work by
the assumption of diagonal damping. Unlike prior time-varying designs, the
time-invariant controller yields smooth, non-oscillatory trajectories---an
important practical advantage for docking. The designs are validated in
simulation on the Cybership\,I hull model.
\end{abstract}

\end{frontmatter}

%=====================================================================
\section{Introduction}

This is a two-part paper in which Part I covers stabilization/parking for
ships, while Part II \citep{partII} deals with tracking a leader ship and
docking to it. The two problems differ substantially and their solutions,
necessarily, employ significantly different techniques in design and analysis.

\paragraph{Problem.}
A surface vessel with two main thrusters and no side thruster
has a surge force and a yaw moment as inputs and no input in sway. Stabilizing
its position and heading at a fixed pose is the dynamic positioning problem for
this underactuated vessel \citep{mazenc,pettersen04}. The vessel does not satisfy
Brockett's necessary condition \citep{brockett}, so no continuous time-invariant
feedback stabilizes it in Cartesian coordinates \citep{pettersen95}. In the existing literature, two strategies have been chosen to circumvent this obstruction: time-varying feedback and discontinuous switching between time-invariant feedback laws. In this paper we propose a third strategy: a smooth time-invariant feedback in polar coordinates.

Time-varying feedback has produced global results for the underactuated ASV. \cite{mazenc} proved global asymptotic stability (GAS); \cite{dongguo} and \cite{ma09} gave further GAS designs by periodic time-varying feedback; all these GAS designs assume diagonal inertia and
damping matrices. Moreover, they produce persistent oscillations in position and heading. For related time-varying and discontinuous results, see also \cite{reyhanoglu99,jiang02,do02,do03,pan26}. For the same class of vessel models, source seeking without
position measurements is achieved in \citet{wang23}.

Three gaps motivate this paper. \emph{First}, no prior design for the underactuated ASV is derived from a polar-coordinate CLF. \emph{Second}, every GAS design uses time-varying feedback; \citeauthor{pan26}’s discontinuous design does not establish global asymptotic stability of the full state. \emph{Third}, all GAS designs require both diagonal inertia and diagonal damping; relaxing either assumption yields at best practical stability~\citep{luo21,zhang22}. These gaps define the contributions below.

The motivation for the proposed approach is that for the unicycle, polar coordinates remove the obstruction established by \cite{brockett}: the target $\rho=0$ is a singularity of the polar chart, so global asymptotic stability in polar variables does not require the equilibrium to be a smooth fixed point of the vector field, and the Brockett condition does not apply in the usual form.
{Time-invariant feedback in polar coordinates was pioneered for the unicycle by \citet{aicardi95}.} \citet{part1} gives time-invariant feedback laws with strict control Lyapunov
functions (CLFs), designed separately for the distance and for the steering
angles, and \citet{part2} adds inverse-optimal, adaptive and safe designs. The
approach has been extended to the tractor-trailer \citep{tractortrailer} and to a
3D vehicle \citep{threeD}. These are driftless kinematic models, in which the
vehicle moves along its heading. A surface vessel slips sideways: its sway speed
is not actuated and evolves with its own damping. This paper extends the polar
approach to this vessel, first with the surge speed and the yaw rate as inputs
and then with the surge force and the yaw moment.

\paragraph{Contributions.}\leavevmode\par\nobreak\vspace{-\parskip}
\begingroup\topsep=0pt\partopsep=0pt
\begin{itemize}
\item A zero-sideslip-point construction that reduces the underactuated surface
vessel to the unicycle of~\citet{part1}: a point whose position relative to the
hull---determined by the surge and sway speeds---moves along the vessel's
heading, so the slipping vessel is steered as the unicycle. This reduction is a
methodological bridge: unicycle control laws, with appropriate analysis, become
candidates for the surface vessel. 
\item Time-invariant feedback that globally stabilizes the underactuated surface vessel
in polar coordinates, with the surge speed and yaw rate as inputs and, by one
step of backstepping, with the surge force and yaw moment, yielding smooth,
non-oscillatory vessel trajectories---an important practical advantage for
docking, in contrast to the persistent oscillations inherent in the periodic time-varying designs of \citet{mazenc,dongguo}.
\item Stabilization of hulls with fore-aft asymmetric damping, excluded in
\citet{mazenc,dongguo} by the assumption of diagonal damping, by a Lyapunov
function in which an energy of the surge and sway speeds dominates the coupling
between yaw rate and sway that this asymmetry creates.
\end{itemize}\endgroup
Each problem is solved with two steering laws of \citet{part1}, one that
allows assignment of the local eigenvalues and one that keeps the desired
position from being directly behind the vessel. The hull condition excludes a
coupling of surge and sway speeds in the sway damping (Assumption
\ref{as:coeff}, Remark \ref{rem:Y1}). The estimates are in polar variables;
in Cartesian variables they depend also on the initial polar angle
(Remark \ref{rem:sense}). For the model with the surge speed and yaw rate as
inputs, the feedback has one integrator.

%=====================================================================
\paragraph{Nomenclature.}\leavevmode\par
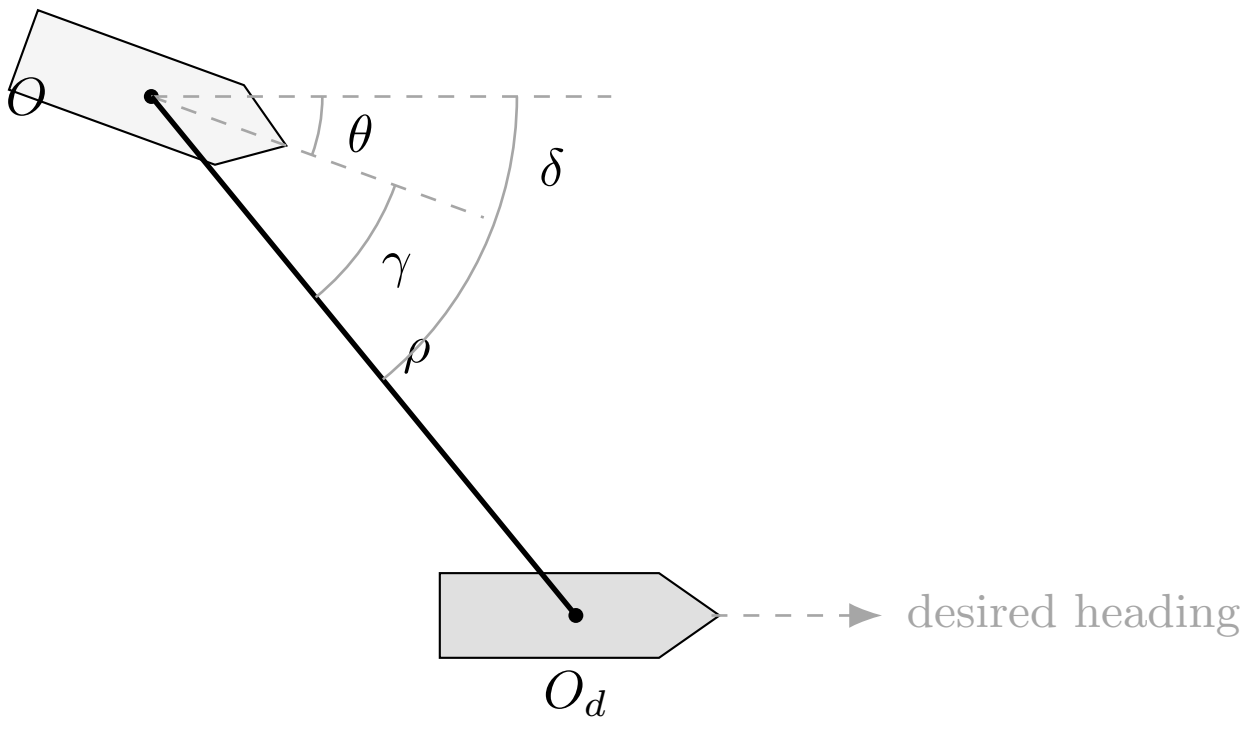
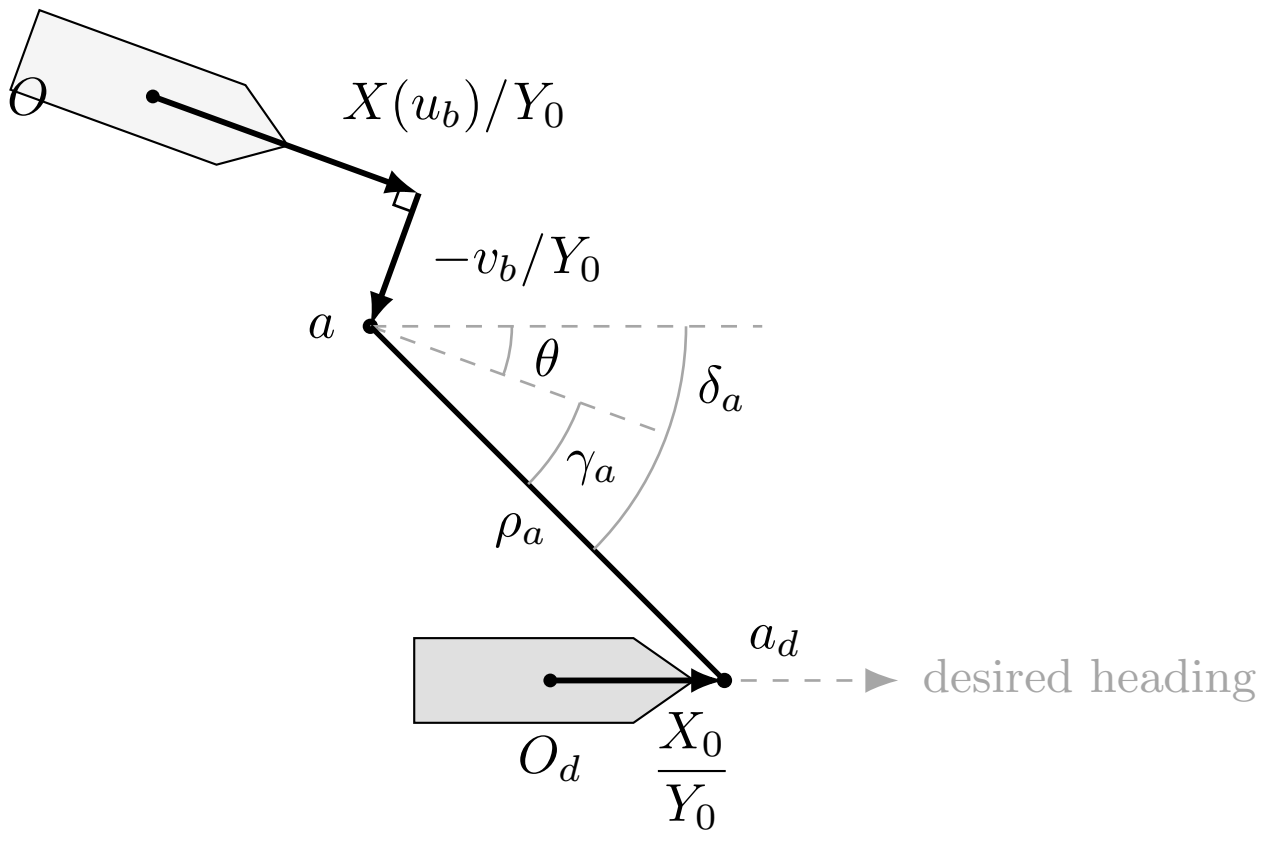
\begin{figure*}[t]
\centering
\begin{subfigure}[t]{0.49\textwidth}
\centering
\begin{tikzpicture}[scale=0.78]
\coordinate (T) at (0,0);
\coordinate (P) at (-3.6,4.4);
\pic at (T) {hull=black!12};
\fill (T) circle (1.8pt);
\node[below=7pt] at (T) {$O_d$};
\draw[ref,->] (1.15,0) -- (2.6,0) node[right=1pt] {\small desired heading};
\pic[rotate=-20] at (P) {hull=black!4};
\fill (P) circle (1.8pt);
\node[left=16pt] at (P) {$O$};
\draw[ref] (P) -- ++(0:3.9);
\draw[ref] (P) -- ++(-20:3.0);
\draw[line width=1.0pt] (P) -- (T);
\node at (-1.35,2.20) {$\rho$};
\draw[ang] (P) ++(0:1.45) arc (0:-20:1.45);
\node at ($(P)+(-10:1.80)$) {$\theta$};
\draw[ang] (P) ++(-20:2.20) arc (-20:-50.7:2.20);
\node at ($(P)+(-35.3:2.55)$) {$\gamma$};
\draw[ang] (P) ++(0:3.10) arc (0:-50.7:3.10);
\node at ($(P)+(-10:3.45)$) {$\delta$};
\end{tikzpicture}
\caption{Polar variables of the vessel. $O$ is the body-frame origin and $O_d$
its desired position. The dashed rays through $O$ are parallel to the desired
heading and to the vessel's heading. All angles are positive as drawn.}
\label{fig:polar}
\end{subfigure}\hfill
\begin{subfigure}[t]{0.49\textwidth}
\centering
\begin{tikzpicture}[scale=0.72]
\coordinate (Od) at (0,0);
\coordinate (Ad) at (1.6,0);
\coordinate (O) at (-3.651,5.364);
\coordinate (C) at (-1.208,4.475);
\coordinate (A) at (-1.653,3.253);
% desired pose
\pic at (Od) {hull=black!12};
\fill (Od) circle (1.8pt);
\node[below=7pt] at (Od) {$O_d$};
\draw[leg] (Od) -- (Ad);
\fill (Ad) circle (2.0pt);
\node[above right=1pt] at (Ad) {$a_d$};
\node[below=2pt] at (1.30,-0.02) {$\dfrac{X_0}{Y_0}$};
\draw[ref,->] (1.75,0) -- (3.2,0) node[right=1pt] {\small desired heading};
% vessel
\pic[rotate=-20] at (O) {hull=black!4};
\fill (O) circle (1.8pt);
\node[left=16pt] at (O) {$O$};
\draw[leg] (O) -- (C);
\node[above right=1pt] at ($(O)!0.62!(C)$) {$X(u_b)/Y_0$};
\draw[leg] (C) -- (A) node[midway,right=3pt] {$-v_b/Y_0$};
\draw[line width=0.5pt] ($(C)+(-0.169,0.062)$) -- ($(C)+(-0.231,-0.107)$) -- ($(C)+(-0.062,-0.169)$);
\fill (A) circle (2.0pt);
\node[left=3pt] at (A) {$a$};
% rays and range
\draw[ref] (A) -- ++(0:3.6);
\draw[ref] (A) -- ++(-20:2.9);
\draw[line width=1.0pt] (A) -- (Ad);
\node at (-0.27,1.38) {$\rho_a$};
\draw[ang] (A) ++(0:1.30) arc (0:-20:1.30);
\node at ($(A)+(-10:1.65)$) {$\theta$};
\draw[ang] (A) ++(-20:2.05) arc (-20:-45:2.05);
\node at ($(A)+(-32.5:2.42)$) {$\gamma_a$};
\draw[ang] (A) ++(0:2.90) arc (0:-45:2.90);
\node at ($(A)+(-10:3.28)$) {$\delta_a$};
\end{tikzpicture}
\caption{The zero-sideslip point $a$ and its polar variables. Starting at the body-frame
origin $O$, the first arrow runs along the forward axis of the vessel with
length $X(u_b)/Y_0$, and the second arrow runs at a right angle to it, along
the starboard axis, with length $-v_b/Y_0$; it ends at $a$. Starting at the
desired position $O_d$, the arrow of length $X_0/Y_0$ along the desired heading
ends at $a_d$. The dashed rays through $a$ are parallel to the desired heading
and to the vessel's heading. Drawn for positive lengths.}
\label{fig:a}
\end{subfigure}
\caption{Polar variables: (a) of the body-frame origin $O$; (b) of the zero-sideslip point $a$.}
\label{fig:polarvars}
\end{figure*}

\begingroup\parskip=0pt\interlinepenalty=10000\relax
\nomgroup{.8pt}{Polar variables of the vessel (Fig.~\ref{fig:polar})}
\nomitem{O,\ O_d}{Body-frame origin of the vessel; its desired position.}
\nomitem{\rho}{Distance from $O$ to $O_d$.}
\nomitem{\delta}{Polar angle: angle from the desired heading to the line from $O$ to $O_d$.}
\nomitem{\gamma}{Line-of-sight angle: angle from the vessel's heading to the line from $O$ to $O_d$.}
\nomitem{\theta}{Heading error: angle from the desired heading to the vessel's heading; $\theta=\delta-\gamma$ where $\rho>0$.}
\nomgroup{.5pt}{Speeds and inputs}
\nomitem{u_b,\ v_b,\ r_b}{Surge (forward) speed, sway (starboard) speed and yaw rate of the vessel.}
\nomitem{\tau_u,\ \tau_r}{Mass-normalized surge force and yaw moment.}
\nomgroup{.5pt}{Hull coefficients}
\nomitem{X(u_b)}{Sway acceleration per unit yaw rate, $X(u_b)=X_0+X_1u_b$: turning at yaw rate $r_b$ pushes the hull sideways with acceleration $X(u_b)r_b$ in \eqref{eq:p-v}.}
\nomitem{X_0,\ X_1}{Hull coefficients of $X(u_b)$, $X_1<0$.}
\nomitem{Y_0}{Sway damping rate, $Y_0<0$.}
\nomitem{f_u,\ f_r}{Unforced surge and yaw accelerations.}
\nomgroup{.5pt}{Zero-sideslip point $a$ and its polar variables (Fig.~\ref{fig:a})}
\nomitem{a}{Zero-sideslip point (Definition \ref{def:a}): its velocity points along the vessel's heading. Displaced from $O$ by $X(u_b)/Y_0$ along the forward axis and by $-v_b/Y_0$ along the starboard axis.}
\nomitem{a_d}{Position of $a$ when the vessel is at rest at the desired pose: displaced from $O_d$ by $X_0/Y_0$ along the desired heading.}
\nomitem{\rho_a}{Distance from $a$ to $a_d$.}
\nomitem{\delta_a}{Polar angle of $a$: angle from the desired heading to the line from $a$ to $a_d$.}
\nomitem{\gamma_a}{Line-of-sight angle of $a$: angle from the vessel's heading to the line from $a$ to $a_d$.}
\nomitem{u_a}{Forward speed of $a$. The lateral speed of $a$ is zero.}
\nomgroup{.5pt}{Design}
\nomitem{k_1,\dots,k_5}{Positive gains.}
\nomitem{\omega^\star}{Steering law of \citet[Thm.~7]{part1}, evaluated at $(\delta_a,\gamma_a)$.}
\nomitem{z}{Backstepping variable of \citet[Thm.~7]{part1}, $z=\gamma_a+\tfrac12\arctan(2k_2\delta_a)$.}
\nomitem{\varsigma}{Yaw-rate error, $\varsigma=r_b-\omega^\star$ (six-state design only).}
\nomitem{V,\ V_6}{Strict control Lyapunov functions of the closed loops of Theorems \ref{thm:four} and \ref{thm:six}.}
\par\vskip3pt\hrule height .8pt\vskip6pt
\endgroup

%=====================================================================
\section{Polar model}

The model is the Cartesian model \eqref{eq:cart-kin}, \eqref{eq:cart-dyn} of
Appendix \ref{app:cart}, written in the polar variables \eqref{eq:cart-error}, \eqref{eq:cart-polar}
and with $Y(u_b)=Y_0$ and
$X(u_b)=X_0+X_1u_b$, the sway acceleration per unit yaw rate:
\begin{subequations}
\begin{align}
\dot\rho &= -u_b\cos\gamma - v_b\sin\gamma, \label{eq:p-rho}\\
\dot\delta &= \frac{u_b\sin\gamma - v_b\cos\gamma}{\rho}, \label{eq:p-delta}\\
\dot\gamma &= \frac{u_b\sin\gamma - v_b\cos\gamma}{\rho} - r_b, \label{eq:p-gamma}\\
\dot v_b &= X(u_b)\, r_b + Y_0\, v_b, \label{eq:p-v}\\
\dot u_b &= f_u(u_b,v_b,r_b) + \tau_u, \label{eq:udot}\\
\dot r_b &= f_r(u_b,v_b,r_b) + \tau_r. \label{eq:rdot}
\end{align}
\end{subequations}
The origin of the body-fixed frame, $O$, is the pivot point of the vessel (Appendix~\ref{app:cart}), the centerline point at which the actuators produce no sway acceleration, which is why no input appears in~\eqref{eq:p-v}.

The \emph{four-state model} is \eqref{eq:p-rho}--\eqref{eq:p-v}: state
$(\rho,\delta,\gamma,v_b)$, input $(u_b,r_b)$. The \emph{six-state model} is
\eqref{eq:p-rho}--\eqref{eq:rdot}: state $(\rho,\delta,\gamma,v_b,u_b,r_b)$,
input $(\tau_u,\tau_r)$. The six-state model is the four-state model with one
integrator added at each input. With $v_b\equiv0$, equations
\eqref{eq:p-rho}--\eqref{eq:p-gamma} are the unicycle \citep[Eq.~(2)]{part1}.

\begin{assumption}
$Y(u_b)=Y_0$ for all $u_b$, with $Y_0<0$, and $X_1<0$.
\label{as:coeff}
\end{assumption}

\begin{remark}[Validity of Assumption~\ref{as:coeff}]\label{rem:coeff}
Assumption \ref{as:coeff} holds for the coefficients of
\eqref{eq:Xu}, \eqref{eq:Yu} when $m_{23}=0$ or $m_{11}=m_{22}$, with arbitrary $d_{23}$ and
$d_{32}$. For $m_{23}=0$:
\begin{equation}
\frac{X_0}{Y_0} = \frac{d_{23}}{d_{22}}, \qquad
\frac{X_1}{Y_0} = \frac{m_{11}}{d_{22}}, \qquad
-\frac{1}{Y_0} = \frac{m_{22}}{d_{22}} .
\label{eq:coeff-m23}
\end{equation}
The sway--yaw added-mass coupling $m_{23}$ is often assumed negligible in control-oriented models of surface
vessel hulls. The condition $m_{11}=m_{22}$
holds when the surge and sway inertia and added mass are equal. The condition $X_1<0$ holds whenever
$m_{11}m_{33}>m_{23}^2$, which follows from positive definiteness of the
mass matrix and is satisfied for all physical hulls. The condition $Y_0<0$
is the standard sway damping condition. Remark~\ref{rem:Y1} addresses the complementary case
$m_{23}(m_{22}-m_{11})\neq 0$.
\end{remark}
\emph{Four-state problem.} For the four-state model, find a feedback for
$(u_b,r_b)$ such that the closed-loop solutions are bounded and
\begin{equation}
\rho\to0, \quad \theta\to0, \quad v_b\to0, \quad u_b\to0, \quad r_b\to0 .
\label{eq:objective}
\end{equation}

\emph{Six-state problem.} For the six-state model, find a feedback for
$(\tau_u,\tau_r)$ such that the closed-loop solutions are bounded,
\eqref{eq:objective} holds and $(\tau_u,\tau_r)\to0$.

Both problems are solved by steering the zero-sideslip point $a$ of
Definition \ref{def:a}, whose position relative to the hull depends on the
surge and sway speeds and whose velocity points along the vessel's heading, to
its desired position $a_d$. The four-state problem is
solved by dynamic feedback of order one (Theorem \ref{thm:four}), and the
six-state problem by static feedback (Theorem \ref{thm:six}).

\emph{Environmental disturbance.}
Environmental forces from wind, ocean currents, or both act on the vessel and
modify the dynamic equations. Following \citet[Sect.~8.1]{fossen11} and
\citet{caharija16}, we model the net environmental force as a horizontal force of magnitude
$F\geq0$ acting at $O$ in a fixed direction at the angle $\varphi$ from the
desired heading. Since $\theta=\delta-\gamma$ is the angle from the desired
heading to the vessel's heading, $\varphi-\theta$ is the angle from the
vessel's heading to the force, and the disturbed sway, surge, and yaw
dynamics are
\begin{subequations}
\begin{align}
\dot v_b &= X(u_b)\,r_b + Y_0\,v_b
           + \tfrac{m_{33}}{\Delta}\,F\sin(\varphi-\theta), \label{eq:vdot-F}\\
\dot u_b &= f_u + \tau_u
           + \tfrac{1}{m_{11}}\,F\cos(\varphi-\theta), \label{eq:udot-F}\\
\dot r_b &= f_r + \tau_r
           - \tfrac{m_{23}}{\Delta}\,F\sin(\varphi-\theta), \label{eq:rdot-F}
\end{align}
\end{subequations}
where $\Delta = m_{22}m_{33}-m_{23}^2>0$. 
In the case $m_{23}=0$ of Assumption~\ref{as:coeff}, \eqref{eq:rdot-F} reduces to
$\dot r_b = f_r+\tau_r$, so the disturbance enters only the sway and surge
dynamics.

For a surface vessel in wind, the magnitude $F = \tfrac{1}{2}\rho_{\mathrm{air}} V_{rw}^2
c_Y A_{Lw}$, with $\rho_{\mathrm{air}}$ the air density, is proportional to the square of the relative wind speed and to
the lateral wind-force coefficient \citep[Sect.~8.1]{fossen11}; it is
constant for a steady wind. For an irrotational constant ocean current of
speed $V_c$ at inertial angle $\varphi$, the dominant sway disturbance term
in \eqref{eq:vdot-F} has effective magnitude $F\approx d_{22}V_c$
\citep[Sect.~8.3]{fossen11}. 

%=====================================================================
\section{Zero-sideslip point $a$ and its polar variables}

\begin{definition}[Zero-sideslip point]
The \emph{zero-sideslip point} $a$ is the point displaced from the body-frame
origin $O$ by $X(u_b)/Y_0$ along the forward axis and by $-v_b/Y_0$ along the
starboard axis. Its \emph{desired position} $a_d$ is the position of $a$ when the
vessel is at rest at the desired pose: $a_d$ is displaced from $O_d$ by
$X_0/Y_0$ along the desired heading.
\label{def:a}
\end{definition}

The velocity of $a$ points along the vessel's heading at every instant. Its
component along the starboard axis is
\begin{equation}
\begin{aligned}
&v_b + \frac{X(u_b)}{Y_0}\,r_b - \frac{\dot v_b}{Y_0}\\
&\quad= v_b + \frac{X(u_b)}{Y_0}\,r_b - \frac{X(u_b)\,r_b+Y_0v_b}{Y_0} = 0
\end{aligned}
\label{eq:a-lateral}
\end{equation}
by \eqref{eq:p-v}, and its component along the forward axis is $u_a$ in
\eqref{eq:ua} below. Physically: no point fixed on the hull has zero sideways speed for every
motion of the vessel, because turning pushes the hull sideways by the amount $X(u_b)r_b$ in
\eqref{eq:p-v} and this sideways speed then decays at the rate $-Y_0$. The point
$a$ moves with the hull and, in addition, shifts relative to the hull with the
surge speed and the sway speed, by exactly the amounts that cancel the sideways
slip. A wheeled unicycle has no sideways slip, so $a$ moves as a unicycle
(Proposition \ref{prop:a}), and the designs of \citet{part1} apply to $a$. When
the vessel is at rest, $a$ is the point at the signed distance $X_0/Y_0$
from $O$ along the forward axis, which need not lie on the hull. When the vessel is at rest with $\theta=0$,
$a=a_d$ holds if and only if $O=O_d$.

The polar
variables $(\rho_a,\delta_a,\gamma_a)$ of $a$ relative to $a_d$ have the same
meaning as $(\rho,\delta,\gamma)$ in Fig.~\ref{fig:polar}, with $a$ in place of
$O$ and $a_d$ in place of $O_d$. In terms of $(\rho,\delta,\gamma,v_b)$ and
$u_b$ they are given by
\begin{subequations}
\begin{align}
\rho_a\cos\gamma_a &= \rho\cos\gamma - \frac{X_1}{Y_0}\,u_b
 - \frac{X_0}{Y_0}\bigl(1-\cos(\delta-\gamma)\bigr), \label{eq:F}\\
\rho_a\sin\gamma_a &= \rho\sin\gamma + \frac{1}{Y_0}\,v_b
 - \frac{X_0}{Y_0}\sin(\delta-\gamma), \label{eq:G}\\
\delta_a &= \gamma_a + \delta - \gamma , \label{eq:delta-a}
\end{align}
\end{subequations}
with $\rho_a>0$. The heading error is the same for both sets of variables,
$\delta_a-\gamma_a=\delta-\gamma$.
{With $\xi$ and $\eta$ the right sides of \eqref{eq:F} and \eqref{eq:G},
explicitly,
\begin{equation}
\begin{aligned}
&\rho_a=\sqrt{\xi^2+\eta^2},\qquad
\gamma_a=\atantwo(\eta,\xi)\ (\mathrm{mod}\ 2\pi),\\
&\delta_a=\gamma_a+\delta-\gamma .
\end{aligned}
\label{eq:a-explicit}
\end{equation}
where $\rho_a=\rho_a(\rho,\delta,\gamma,v_b,u_b)$,
$\delta_a=\delta_a(\rho,\delta,\gamma,v_b,u_b)$,
$\gamma_a=\gamma_a(\rho,\delta,\gamma,v_b,u_b)$, i.e., this is not just a
pose-to-pose change of variables but involves the velocity as well.
The transformation is invertible: with $\theta=\delta_a-\gamma_a$ and
\begin{subequations}
\begin{align}
\tilde\xi &= \rho_a\cos\gamma_a + \frac{X_1}{Y_0}\,u_b
 + \frac{X_0}{Y_0}\bigl(1-\cos\theta\bigr), \label{eq:F-inv}\\
\tilde\eta &= \rho_a\sin\gamma_a - \frac{1}{Y_0}\,v_b
 + \frac{X_0}{Y_0}\sin\theta, \label{eq:G-inv}
\end{align}
for $\rho>0$,
\begin{equation}
\begin{aligned}
&\rho=\sqrt{\tilde\xi^2+\tilde\eta^2},\qquad
\gamma=\atantwo(\tilde\eta,\tilde\xi)\ (\mathrm{mod}\ 2\pi),\\
&\delta=\gamma+\theta ,
\end{aligned}
\label{eq:delta-inv}
\end{equation}
\end{subequations}
and $\tilde\xi=\rho\cos\gamma$, $\tilde\eta=\rho\sin\gamma$.}

The forward speed of $a$ is
\begin{equation}
u_a = u_b + \frac{v_b r_b}{Y_0} + \frac{X_1}{Y_0}\,\dot u_b .
\label{eq:ua}
\end{equation}

\begin{proposition}[The zero-sideslip point moves as a unicycle]
Let Assumption \ref{as:coeff} hold. Along the solutions of
\eqref{eq:p-rho}--\eqref{eq:p-v} with $u_b$ differentiable and $\rho_a>0$,
\begin{subequations}
\begin{align}
\dot\rho_a &= -u_a\cos\gamma_a, \label{eq:a-rho}\\
\dot\delta_a &= \frac{u_a}{\rho_a}\sin\gamma_a, \label{eq:a-delta}\\
\dot\gamma_a &= \frac{u_a}{\rho_a}\sin\gamma_a - r_b, \label{eq:a-gamma}\\
\dot v_b &= Y_0 v_b + (X_0+X_1u_b)\,r_b  \label{eq:a-v}\\
\dot u_b &= \frac{Y_0}{X_1}\bigl(u_a - u_b\bigr) - \frac{1}{X_1}\,v_b r_b,
 \label{eq:a-u}
\end{align}
\end{subequations}
\label{prop:a}
\end{proposition}

The proof is in Appendix \ref{app:proof}. Equations
\eqref{eq:a-rho}--\eqref{eq:a-gamma} are the unicycle \citep[Eq.~(2)]{part1} with
forward speed $u_a$ and steering rate $r_b$; neither $u_b$ nor $v_b$ appears in
them. Equation \eqref{eq:a-u} is \eqref{eq:ua} solved for $\dot u_b$, and
\eqref{eq:a-v} is \eqref{eq:p-v}.

In \eqref{eq:a-rho}--\eqref{eq:a-v} the surge speed $u_b$ is a state, driven
by $u_a$ through \eqref{eq:a-u}, because the position of $a$ depends on $u_b$.
A feedback for $(u_a,r_b)$ that is static in $(\rho_a,\delta_a,\gamma_a,u_b,v_b)$
is therefore, for the four-state model, a dynamic feedback of order one whose
controller state is $u_b$; the closed loop has five states.

%=====================================================================
\section{Roadmap of results}\label{sec:roadmap}
 
{Each problem is solved with two steering laws of \citet{part1}, GloBa
\citep[Thm.~7]{part1} and BoLSA \citep[Thm.~2]{part1} (Table \ref{tab:results}).
The four designs share the zero-sideslip point, the surge law and the energy
term $(X_1^2u_b^2+v_b^2)/(k_1Y_0X_1)$, and differ in the steering law.
Theorems \ref{thm:four} and \ref{cor:bolsa4} solve the four-state problem with
dynamic feedback of order one, because the velocity $u_a$ in \eqref{eq:ua}
contains $\dot u_b$. Theorems \ref{thm:six} and \ref{cor:bolsa} solve the
six-state problem with static feedback, by one step of backstepping on the yaw
rate. BoLSA keeps $|\gamma_a(t)|<\pi$, so the desired position is never directly
behind the vessel; GloBa allows the local eigenvalues to be assigned at any
negative real values (Remark \ref{rem:eig}) and requires no gain condition.}
 
\begin{table*}[t]
\centering
\small
\renewcommand{\arraystretch}{1.25}
\begin{tabular}{@{}>{\raggedright}p{0.7in}>{\raggedright}p{1.4in}>{\raggedright}p{1.7in}>{\raggedright\arraybackslash}p{2.1in}@{}}
\toprule
Result & Steering law & Conditions beyond Assumption \ref{as:coeff} & Property of the steering law\\
\midrule
\multicolumn{4}{@{}l}{\emph{Four-state model, inputs $(u_b,r_b)$: dynamic feedback of order one}}\\
\midrule
Thm.~\ref{thm:four} & GloBa, $\gamma_a\in\mathbb{R}$ & $k_1,\dots,k_4>0$ & Local eigenvalues assignable; no gain condition.\\
Cor.~\ref{cor:X0} & GloBa, $\gamma_a\in\mathbb{R}$ & $X_0=0$ & As Thm.~\ref{thm:four}; $a_d=O_d$.\\
Thm.~\ref{cor:bolsa4} & BoLSA, $|\gamma_a|<\pi$ & $k_1k_3\ge k_2^2$ & $|\gamma_a(t)|<\pi$: desired position never directly behind the vessel.\\
\midrule
\multicolumn{4}{@{}l}{\emph{Six-state model, inputs $(\tau_u,\tau_r)$: static feedback, one step of backstepping on $r_b$}}\\
\midrule
Thm.~\ref{thm:six} & GloBa, $\gamma_a\in\mathbb{R}$ & $k_1,\dots,k_5>0$ & As Thm.~\ref{thm:four}.\\
Thm.~\ref{cor:bolsa} & BoLSA, $|\gamma_a|<\pi$ & $k_1k_3\ge k_2^2$; $k_5,c>0$ & As Thm.~\ref{cor:bolsa4}; simulated in Fig.~\ref{fig:fan}.\\
\bottomrule
\end{tabular}
\caption{Results by model and steering law.}
\label{tab:results}
\end{table*}
 
\begin{remark}[Backstepping in related designs]
The unicycle design \citep{part1} takes one backstepping step, from the heading to the angular velocity. The vessel needs a second step because of the yaw inertia. The tractor-trailer \citep{tractortrailer} has no inertia, yet it also needs two steps, from the trailer heading through the hitch angle. The 3D vehicle \citep{threeD} takes one step in each of yaw and pitch.
\label{rem:difficulty}
\end{remark}

\section{Dynamic feedback for four-state model}

\begin{theorem}[Four-state problem solved by GloBa with dynamic feedback of order one]
Let Assumption \ref{as:coeff} hold and let $k_1,k_2,k_3,k_4>0$. Consider the
four-state model \eqref{eq:p-rho}--\eqref{eq:p-v} under the dynamic feedback
\begin{subequations}
\begin{align}
\dot u_b &= \frac{Y_0}{X_1}\bigl(u_a-u_b\bigr) - \frac{1}{X_1}\,v_b r_b,
 \notag\\
 &\qquad u_a = k_1\rho_a\cos\gamma_a, \label{eq:fb-ua}\\
r_b &= \omega^\star(\delta_a,\gamma_a)
 = \frac{k_1}{2}\sin(2\gamma_a) + k_4 z \notag\\
 &\quad + \frac{k_1}{2}\,\frac{k_2}{1+4k_2^2\delta_a^2}\,\sin(2\gamma_a)
 + k_3\,\psi(z,\gamma_a)\,\delta_a . \label{eq:fb-r}
\end{align}
\end{subequations}
where the controller state $u_b$ is also the surge-speed input of the
four-state model, $(\rho_a,\delta_a,\gamma_a)$ are the functions of
$(\rho,\delta,\gamma,v_b,u_b)$ given by \eqref{eq:F}--\eqref{eq:delta-a}, and
\begin{subequations}
\begin{align}
z &= \gamma_a + \tfrac12\arctan(2k_2\delta_a), \label{eq:z}\\
\psi(z,\gamma) &= \frac{\sin z}{z}\,\cos(z-2\gamma), \qquad \psi(0,\gamma)=\cos(2\gamma). \label{eq:psi}
\end{align}
\end{subequations}
%
% with controller state $u_b$, alternatively written, for $u_b(0)=0$, as
% %
% \begin{equation}
% u_b(t)=\frac{1}{(X_1/Y_0)\,s+1}\Bigl[u_a(t)-\frac{v_b(t)\,r_b(t)}{Y_0}\Bigr],
% \label{eq:ub-filter}
% \end{equation}
% %
The function
\begin{equation}
V(\rho_a,\delta_a,\gamma_a,u_b,v_b)
 = \rho_a^2 + \frac{X_1^2u_b^2+v_b^2}{k_1Y_0X_1}
 + \Phi\Bigl(\delta_a^2+\frac{k_1}{k_3}\,z^2\Bigr)
\label{eq:V}
\end{equation}
where
\begin{subequations}
\begin{align}
\Phi(s) &= s + \frac{\kappa}{6k_2^2}\Bigl[\bigl(1+4k_2^2s\bigr)^{3/2}-1\Bigr],
 \label{eq:Phi}\\
\kappa &= \frac{X_0^2}{k_1Y_0^2|X_1|}
 \max\Bigl\{\frac{\bigl(k_3+k_1k_2(1+k_2)\bigr)^2}{k_1k_2}, \notag\\
 &\qquad\qquad\qquad \frac{k_3\bigl(k_4+k_1(1+k_2)\bigr)^2}{k_1k_4}\Bigr\} , \label{eq:kappa}
\end{align}
\end{subequations}
is positive definite and radially unbounded in its five arguments, and along the
closed-loop solutions
\begin{equation}
\begin{aligned}
\dot V \le{}& -k_1\rho_a^2\cos^2\gamma_a - \frac{u_b^2}{k_1}
 - \frac{v_b^2}{k_1|X_1|}\\
& - \frac{2k_1k_2\,\delta_a^2}{\sqrt{1+4k_2^2\delta_a^2}}
 - \frac{2k_1k_4}{k_3}\,z^2 .
\end{aligned}
\label{eq:Vdot}
\end{equation}
For every initial condition with $\rho_a(0)>0$, the solution exists and is
unique for all $t\ge0$, satisfies $\rho_a(t)>0$, and
\begin{equation}
\Upsilon(t) \le \beta\bigl(\Upsilon(0),t\bigr), \qquad
\Upsilon = \rho_a + |\delta_a| + |\gamma_a| + |u_b| + |v_b| ,
\label{eq:KL}
\end{equation}
for a class $\mathcal{KL}$ function $\beta$ that does not depend on the initial
condition. In particular, \eqref{eq:objective} holds.
\label{thm:four}
\end{theorem}

The closed-loop state is $(\rho_a,\delta_a,\gamma_a,u_b,v_b)$. Equations
\eqref{eq:p-rho}--\eqref{eq:p-gamma} describe the vessel on intervals where
$\rho>0$; passage of $O$ through $O_d$ does not interrupt the closed-loop
solution.

{
\begin{remark}[Sense of stability]\label{rem:sense}
The estimate \eqref{eq:KL} is a global uniform asymptotic stability estimate in
the polar variables of $a$: $\beta$ does not depend on the initial condition.
In the polar variables of $O$, \eqref{eq:F}, \eqref{eq:G}, \eqref{eq:F-inv},
\eqref{eq:G-inv} and $\tilde\xi=\rho\cos\gamma$, $\tilde\eta=\rho\sin\gamma$ give
$|\rho-\rho_a|\le|X_1/Y_0||u_b|+|v_b|/|Y_0|+|X_0/Y_0||\theta|$. With
$|\theta|\le|\delta_a|+|\gamma_a|$ and $|\delta_a|\le|\gamma_a|+|\theta|$,
\eqref{eq:KL} gives
\begin{equation}
\begin{aligned}
&\rho(t)+|\theta(t)|+|u_b(t)|+|v_b(t)|\\
&\le c\,\beta\Bigl(c\bigl(\rho(0)+|\theta(0)|+|u_b(0)|+|v_b(0)|\bigr)\\
&\qquad\qquad +2|\gamma_a(0)|,\,t\Bigr),
\end{aligned}
\label{eq:KL-O}
\end{equation}
with $c=1+\max\{|X_1|,1,|X_0|\}/|Y_0|$. The term $2|\gamma_a(0)|$, at most
$2\pi$ when $\gamma_a(0)$ is taken in $(-\pi,\pi]$, does not vanish with the
initial error of the hull. With it, \eqref{eq:KL-O} expresses \emph{global uniform
Lagrange asymptotic stability} (GULAS) of the desired pose, as in
\citet[Cor.~2]{part1}: global uniform boundedness and global uniform
attractivity. The bearing $\delta$ and the
line-of-sight angle $\gamma=\delta-\theta$ of $O$ are not bounded by
\eqref{eq:KL-O}; their behavior is described by Proposition \ref{prop:dir}
below.
\end{remark}
}

The steering law \eqref{eq:fb-r} and the reference speed $u_a = k_1\rho_a\cos\gamma_a$ are the GloBa design of \citet[Thm.~7]{part1}, applied at the zero-sideslip coordinates $(\rho_a,\delta_a,\gamma_a)$. The CLF $V$ is new: the energy term $(X_1^2u_b^2+v_b^2)/(k_1Y_0X_1)$ captures the surge and sway dynamics, and the weight $\Phi$ in \eqref{eq:Phi} modifies the angular part to absorb the cross-coupling that arises when $X_0\ne0$. The gain $\kappa$ in \eqref{eq:kappa} is derived from the proof. For $X_0=0$ one has $\kappa=0$ and $\Phi(s)=s$, so $V$ reduces to a quadratic CLF.

\begin{proof}[Proof of Theorem \ref{thm:four}]
\emph{Closed loop.} By Proposition \ref{prop:a}, the closed loop is
\eqref{eq:a-rho}--\eqref{eq:a-v}, and by \eqref{eq:a-u} and the controller equation
of \eqref{eq:fb-ua}, the forward speed of $a$ is $u_a=k_1\rho_a\cos\gamma_a$.

\emph{Range and angles.} With \eqref{eq:fb-ua} and \eqref{eq:fb-r}, equations
\eqref{eq:a-rho}--\eqref{eq:a-gamma} are the closed loop of
\citet[Thm.~7]{part1}. Hence, with
$V_{\delta\gamma}=\delta_a^2+(k_1/k_3)z^2$ and by \citet[Eqs.~(5),~(115)]{part1},
\begin{subequations}
\begin{align}
\frac{d}{dt}\,\rho_a^2 &= -2k_1\rho_a^2\cos^2\gamma_a , \label{eq:rho-dot}\\
\dot V_{\delta\gamma} &= -W, \qquad
W = \frac{2k_1k_2\,\delta_a^2}{\sqrt{1+4k_2^2\delta_a^2}}
 + \frac{2k_1k_4}{k_3}\,z^2 . \label{eq:W}
\end{align}
\end{subequations}

\emph{Speeds.} Multiplying \eqref{eq:a-u} by $X_1^2u_b$ and \eqref{eq:a-v} by
$v_b$ and adding, the two terms in $u_bv_br_b$ cancel:
\begin{equation}
\begin{aligned}
\frac{d}{dt}\,\frac{X_1^2u_b^2+v_b^2}{2}
 ={}& -Y_0X_1\,u_b^2 + Y_0\,v_b^2\\
 & + Y_0X_1\,u_b\,u_a + X_0\,v_b r_b .
\end{aligned}
\label{eq:energy}
\end{equation}
Since $Y_0X_1>0$ and $Y_0<0$, Young's inequality applied to the last two terms
of \eqref{eq:energy}, with $u_a=k_1\rho_a\cos\gamma_a$, gives
\begin{equation}
\begin{aligned}
\frac{d}{dt}\,\frac{X_1^2u_b^2+v_b^2}{k_1Y_0X_1}
 \le{}& -\frac{u_b^2}{k_1} - \frac{v_b^2}{k_1|X_1|}\\
 & + k_1\rho_a^2\cos^2\gamma_a + \frac{X_0^2}{k_1Y_0^2|X_1|}\,r_b^2 .
\end{aligned}
\label{eq:E-bound}
\end{equation}
The third term on the right of \eqref{eq:E-bound} is half of the right side of
\eqref{eq:rho-dot} with the opposite sign.

\emph{Yaw rate.} Since $\psi(z,\gamma)=\cos(z-2\gamma)\sin(z)/z$, one has
$|\psi|\le1$. With $|\sin(2\gamma_a)|\le2|\gamma_a|$ and
$|\gamma_a|\le|z|+k_2|\delta_a|$, \eqref{eq:fb-r} gives
$|r_b|\le\bigl(k_4+k_1(1+k_2)\bigr)|z|+\bigl(k_3+k_1k_2(1+k_2)\bigr)|\delta_a|$.
Since $\sqrt{1+4k_2^2\delta_a^2}\le\sqrt{1+4k_2^2V_{\delta\gamma}}$, by
\eqref{eq:W} and \eqref{eq:kappa},
\begin{equation}
\frac{X_0^2}{k_1Y_0^2|X_1|}\,r_b^2
 \le \kappa\sqrt{1+4k_2^2V_{\delta\gamma}}\;W .
\label{eq:r-bound}
\end{equation}

\emph{Sum.} By \eqref{eq:Phi}, $\Phi'(s)=1+\kappa\sqrt{1+4k_2^2s}$. Adding
\eqref{eq:rho-dot}, \eqref{eq:E-bound} and
$\Phi'(V_{\delta\gamma})\dot V_{\delta\gamma}$, and using \eqref{eq:r-bound},
gives \eqref{eq:Vdot}. The right side of \eqref{eq:Vdot} is zero only where
$\delta_a=z=0$, hence $\gamma_a=0$, and then only where $\rho_a=u_b=v_b=0$.

\emph{Estimate and objective.} By \eqref{eq:rho-dot},
$\rho_a(t)\ge\rho_a(0)e^{-k_1t}>0$. The closed loop is locally Lipschitz and $V$
is nonincreasing, so the solution exists and is unique for all $t\ge0$. The
estimate \eqref{eq:KL} follows from \eqref{eq:Vdot} as in
\citet[proof of Thm.~1]{part1}. By \eqref{eq:KL}, $\rho_a$, $\delta_a$,
$\gamma_a$, $u_b$, $v_b$ converge to zero; then \eqref{eq:F}, \eqref{eq:G} give
$\rho\to0$, $\theta=\delta_a-\gamma_a\to0$, and \eqref{eq:fb-r}
gives $r_b\to0$.
\end{proof}

{
\begin{proposition}[Direction of arrival of the hull]\label{prop:dir}
Let the conditions of Theorem \ref{thm:four} hold with $X_0=0$ and
$k_1<\min\{-Y_0,\,Y_0/X_1\}$. Then along every solution $\rho(t)>0$ for all
sufficiently large $t$, and $\delta(t)\to0$ and $\gamma(t)\to0$ modulo $2\pi$.
\end{proposition}

\begin{proof}[Proof of Proposition \ref{prop:dir}]
Let $\lambda=Y_0/X_1>0$, $p=u_b/\rho_a$ and $q=v_b/\rho_a$. By \eqref{eq:a-rho}
and \eqref{eq:fb-ua}, $\dot\rho_a=-k_1\rho_a\cos^2\gamma_a\ge-k_1\rho_a$, so
$\rho_a(s)\le e^{k_1(t-s)}\rho_a(t)$ for $s\le t$. By Theorem \ref{thm:four},
$\gamma_a\to0$ and $r_b\to0$; let $T$ be such that $u_a\ge0$ on $[T,\infty)$, and
let $\varepsilon=\sup_{s\ge T}|r_b(s)|$. Solving \eqref{eq:fb-ua} and
\eqref{eq:p-v} with $X_0=0$ on $[T,t]$, dividing by $\rho_a(t)$, and using
$0\le u_a\le k_1\rho_a$,
\begin{align*}
|p(t)|&\le e^{-(\lambda-k_1)(t-T)}|p(T)|+\frac{k_1\lambda}{\lambda-k_1}
 +\frac{\varepsilon\,\sup_{[T,t]}|q|}{|X_1|(\lambda-k_1)},\\
p(t)&\ge -e^{-(\lambda-k_1)(t-T)}|p(T)|
 -\frac{\varepsilon\,\sup_{[T,t]}|q|}{|X_1|(\lambda-k_1)},\\
|q(t)|&\le e^{-(|Y_0|-k_1)(t-T)}|q(T)|
 +\frac{|X_1|\,\varepsilon\,\sup_{[T,t]}|p|}{|Y_0|-k_1}.
\end{align*}
For $T$ large, $\varepsilon^2<(\lambda-k_1)(|Y_0|-k_1)$, and the first and
third inequalities give $\sup_{[T,\infty)}(|p|+|q|)<\infty$. Since
$\varepsilon\to0$ as $T\to\infty$, the third gives $q\to0$ and the second gives
$\liminf p\ge0$. By \eqref{eq:F-inv}, \eqref{eq:G-inv} with $X_0=0$,
$\rho\cos\gamma/\rho_a=\cos\gamma_a+(X_1/Y_0)\,p$, whose $\liminf$ is at least
$1$ because $X_1/Y_0>0$, and $\rho\sin\gamma/\rho_a=\sin\gamma_a-q/Y_0\to0$.
Hence $\rho\cos\gamma>0$ for large $t$, $\gamma\to0$ modulo $2\pi$, and
$\delta=\gamma+\theta\to0$ modulo $2\pi$ because $\theta\to0$.
\end{proof}

\begin{remark}[Sideways arrival]\label{rem:sideways}
The condition $k_1<-Y_0$ makes the sway decay faster than $a$ approaches
$a_d$, so the hull arrives along its heading, as a unicycle does. When
$k_1>-Y_0$, the hull can instead arrive sideways, like a skier in a hockey
stop: $\rho\to0$ and $\theta\to0$, while $\delta$ and $\gamma$ tend to a common
nonzero value. This is a broadside arrival, as in berthing alongside a pier,
and no unicycle can perform it. For the hull of Section \ref{sec:sim},
$-Y_0=0.28$ and $Y_0/X_1=0.53$. With the gains of Fig.~\ref{fig:fan},
$k_1=0.1$, all maneuvers arrive with $|\delta|<1^\circ$ when $\rho=1$\,mm.
With $k_1=0.5$, the starts at $45^\circ$ and $315^\circ$ arrive with $\delta$
within $3^\circ$ of $-90^\circ$.
\end{remark}
}

\begin{corollary}[Four-state problem with $X_0=0$ solved by GloBa]
Let Assumption \ref{as:coeff} hold with $X_0=0$, and let $k_1,k_2,k_3,k_4>0$. Then
$a_d=O_d$, the polar variables of $a$ are given by
\begin{subequations}
\begin{align}
\rho_a\cos\gamma_a &= \rho\cos\gamma - \frac{X_1}{Y_0}\,u_b, \label{eq:F0x}\\
\rho_a\sin\gamma_a &= \rho\sin\gamma + \frac{1}{Y_0}\,v_b, \label{eq:G0x}\\
\delta_a &= \gamma_a + \delta - \gamma , \label{eq:delta-a0}
\end{align}
\end{subequations}
and under the dynamic feedback \eqref{eq:fb-ua}, \eqref{eq:fb-r} the function
\begin{equation}
V_0 = \rho_a^2 + \frac{X_1^2u_b^2+v_b^2}{k_1Y_0X_1} + \delta_a^2
 + \frac{k_1}{k_3}\Bigl(\gamma_a+\tfrac12\arctan(2k_2\delta_a)\Bigr)^2
\label{eq:V0}
\end{equation}
satisfies \eqref{eq:Vdot}, and the conclusions of Theorem \ref{thm:four} hold.
\label{cor:X0}
\end{corollary}

\begin{proof}[Proof of Corollary \ref{cor:X0}]
With $X_0=0$, \eqref{eq:F}--\eqref{eq:delta-a} become
\eqref{eq:F0x}--\eqref{eq:delta-a0}, \eqref{eq:kappa} gives $\kappa=0$, and
\eqref{eq:Phi} gives $\Phi(s)=s$, so \eqref{eq:V} is \eqref{eq:V0}.
\end{proof}

For $m_{23}=0$, by \eqref{eq:coeff-m23}, $X_0=0$ holds when $d_{23}=0$, and $a$
is displaced from $O$ by $(m_{11}/d_{22})\,u_b$ along the forward axis and by
$(m_{22}/d_{22})\,v_b$ along the starboard axis. The integrator \eqref{eq:fb-ua}
remains, because the forward displacement depends on $u_b$.

\begin{theorem}[Four-state problem solved by BoLSA]
Let Assumption \ref{as:coeff} hold, and let $k_1,k_2,k_3>0$ with
$k_1k_3\ge k_2^2$. Consider the four-state model
\eqref{eq:p-rho}--\eqref{eq:p-v} under the dynamic feedback consisting of the
controller equation of \eqref{eq:fb-ua} and
\begin{equation}
\begin{aligned}
r_b = \omega^\circ(\delta_a,\gamma_a) ={}& \frac{k_1}{2}\sin(2\gamma_a) + k_2\sin\gamma_a\\
 & + \frac{k_3\cos\gamma_a}{\bigl(1+\tan^2\frac{\gamma_a}{2}\bigr)^2}\,\delta_a ,
\end{aligned}
\label{eq:omega-B}
\end{equation}
with $\gamma_a\in(-\pi,\pi)$ and $\delta_a\in\mathbb{R}$. The function
\begin{equation}
V_4 = \rho_a^2 + \frac{X_1^2u_b^2+v_b^2}{k_1Y_0X_1} + \lambda_4V_{\delta\gamma}
\label{eq:V4B}
\end{equation}
where
\begin{equation}
\begin{aligned}
V_{\delta\gamma} &= k_3\Bigl(1+\frac{2q^2+U}{2qk_2}\Bigr)U
 + \Bigl(\delta_a+2q\tan\frac{\gamma_a}{2}\Bigr)^2,\\
U &= \delta_a^2 + 4q^2\tan^2\frac{\gamma_a}{2}, \qquad q=\sqrt{\frac{k_1}{k_3}} ,
\end{aligned}
\label{eq:Vdg-B}
\end{equation}
\begin{equation}
\begin{aligned}
\lambda_4 &= 1 + \varepsilon\max\Bigl\{
 \frac{2(k_1+k_2)^2+4k_1k_3}{k_1k_2},\; \frac{16k_3^2}{3k_2}\Bigr\},\\
\varepsilon &= \frac{X_0^2}{k_1Y_0^2|X_1|} ,
\end{aligned}
\label{eq:lambda4}
\end{equation}
satisfies, along the closed-loop solutions,
\begin{equation}
\begin{aligned}
\dot V_4 \le{}& -k_1\rho_a^2\cos^2\gamma_a - \frac{u_b^2}{k_1}
 - \frac{v_b^2}{k_1|X_1|}\\
 & - \frac{\lambda_4}{2}\Bigl[8k_1k_2\tan^2\frac{\gamma_a}{2}
 + \frac{3k_2}{2}\Bigl(\delta_a+2q\tan\frac{\gamma_a}{2}\Bigr)^2\Bigr] .
\end{aligned}
\label{eq:V4Bdot}
\end{equation}
For every initial condition with $\rho_a(0)>0$ and $|\gamma_a(0)|<\pi$, the
solution exists and is unique for all $t\ge0$, satisfies $\rho_a(t)>0$ and
$|\gamma_a(t)|<\pi$, and \eqref{eq:KL} holds with $|\gamma_a|$ replaced by
$2|\tan(\gamma_a/2)|$. In particular, \eqref{eq:objective} holds.
\label{cor:bolsa4}
\end{theorem}

\begin{proof}[Proof of Theorem \ref{cor:bolsa4}]
The steering law \eqref{eq:omega-B} is \citet[Eq.~(7)]{part1} with the BoLSA law
\citep[Eq.~(42)]{part1}, and $V_{\delta\gamma}$ is \citet[Eq.~(43)]{part1}. By
Proposition \ref{prop:a} and the controller equation of \eqref{eq:fb-ua}, the
closed loop is \eqref{eq:a-rho}--\eqref{eq:a-v} with
$u_a=k_1\rho_a\cos\gamma_a$, so by \citet[Eqs.~(50),~(57)]{part1}
\begin{equation}
\begin{aligned}
&\dot V_{\delta\gamma} \le -W_4, \\
&W_4 = 8k_1k_2\tan^2\frac{\gamma_a}{2}
 + \frac{3k_2}{2}\Bigl(\delta_a+2q\tan\frac{\gamma_a}{2}\Bigr)^2 .
\end{aligned}
\label{eq:Vdg4dot}
\end{equation}
With $s=\tan(\gamma_a/2)$, one has $|\sin\gamma_a|\le2|s|$ and
$|\cos\gamma_a|/(1+s^2)^2\le1$, so \eqref{eq:omega-B} gives
$|\omega^\circ|\le2(k_1+k_2)|s|+k_3|\delta_a|$, and with
$\delta_a^2\le2(\delta_a+2qs)^2+8q^2s^2$,
\begin{equation}
(\omega^\circ)^2 \le \bigl(8(k_1+k_2)^2+16k_1k_3\bigr)s^2 + 4k_3^2(\delta_a+2qs)^2 .
\label{eq:r-bound-4}
\end{equation}
By \eqref{eq:lambda4}, the last term of \eqref{eq:E-bound}, $\varepsilon r_b^2$, is
then at most $\frac{\lambda_4}{2}W_4$. Adding \eqref{eq:rho-dot},
\eqref{eq:E-bound} and $\lambda_4$ times \eqref{eq:Vdg4dot} gives
\eqref{eq:V4Bdot}. Since $V_{\delta\gamma}\to\infty$ as $|\gamma_a|\to\pi$ and
$V_4$ is nonincreasing, $|\gamma_a(t)|<\pi$. The remaining claims follow as in
the proof of Theorem \ref{thm:four}.
\end{proof}

\begin{remark}[Eigenvalues assignable by GloBa and by BoLSA]
Under Theorem \ref{thm:four} and Theorem \ref{cor:bolsa4}, the
$(\rho_a,\delta_a,\gamma_a)$-loop \eqref{eq:a-rho}--\eqref{eq:a-gamma} with
$u_a=k_1\rho_a\cos\gamma_a$ is the closed loop of \citet{part1}, and
$(u_b,v_b)$ do not enter it. Its linearization at $\rho_a>0$,
$\delta_a=\gamma_a=0$ has the eigenvalue $-k_1$ and the roots of
\begin{subequations}
\begin{align}
&s^2+(k_1k_2+k_4)\,s+k_1(k_2k_4+k_3) &&\text{(GloBa)}, \label{eq:charG}\\
&s^2+k_2\,s+k_1k_3 &&\text{(BoLSA)} , \label{eq:charB}
\end{align}
\end{subequations}
where \eqref{eq:charG} follows from $\dot\delta_a=-k_1k_2\delta_a+k_1z$,
$\dot z=-k_3\delta_a-k_4z$, and \eqref{eq:charB} is \citet[Prop.~2]{part1}.
For any $p_1>0$ and $p_2\ge p_3>0$, the gains $k_1=p_1$,
$k_2=(p_2+p_3/2)/p_1$, $k_3=p_3(2p_2-p_3)/(4p_1)$, $k_4=p_3/2$ place the
eigenvalues of GloBa at $-p_1,-p_2,-p_3$. Under $k_1k_3\ge k_2^2$, the roots of
\eqref{eq:charB} are complex with damping ratio $k_2/(2\sqrt{k_1k_3})\le1/2$.
The remaining eigenvalues of the linearized closed loop are $-Y_0/X_1$ and
$Y_0$, determined by the hull.
\label{rem:eig}
\end{remark}

%=====================================================================
\section{Static feedback for six-state model}

In the six-state model $u_b$ and $r_b$ are states. The surge force $\tau_u$
makes the plant state $u_b$ obey the controller equation of \eqref{eq:fb-ua},
and the yaw moment $\tau_r$ is obtained from \eqref{eq:fb-r} by one step of
backstepping.

\begin{theorem}[Six-state problem solved by GloBa with static feedback]
Let Assumption \ref{as:coeff} hold and let $k_1,\dots,k_5>0$. Consider the
six-state model \eqref{eq:p-rho}--\eqref{eq:rdot} under the static feedback
\begin{subequations}
\begin{align}
\tau_u &= \frac{Y_0}{X_1}\bigl(k_1\rho_a\cos\gamma_a - u_b\bigr)
 - \frac{1}{X_1}\,v_b r_b \notag\\
 &\quad - f_u(u_b,v_b,r_b), \label{eq:tau-u}\\
\tau_r &= - f_r(u_b,v_b,r_b)
 + \frac{k_1}{2}\sin(2\gamma_a)\,\frac{\partial\omega^\star}{\partial\delta_a} \notag\\
 &\quad + \Bigl(\frac{k_1}{2}\sin(2\gamma_a) - r_b\Bigr)
   \frac{\partial\omega^\star}{\partial\gamma_a} \notag\\
 &\quad + \frac{k_1}{k_3}\,z - k_5\bigl(r_b-\omega^\star\bigr), \label{eq:tau-r}
\end{align}
\end{subequations}
with $\omega^\star(\delta_a,\gamma_a)$ and $z$ as in \eqref{eq:fb-r} and
\eqref{eq:z}, and $(\rho_a,\delta_a,\gamma_a)$ given by
\eqref{eq:F}--\eqref{eq:delta-a}. The function
\begin{equation}
\begin{aligned}
&V_6(\rho_a,\delta_a,\gamma_a,u_b,v_b,r_b)
 = \rho_a^2 + \frac{X_1^2u_b^2+v_b^2}{k_1Y_0X_1}\\
&\quad + \Phi_6\Bigl(\delta_a^2+\frac{k_1}{k_3}\,z^2+\bigl(r_b-\omega^\star\bigr)^2\Bigr)
\end{aligned}
\label{eq:V6}
\end{equation}
where $\Phi_6$ is \eqref{eq:Phi} with $\kappa$ replaced by
\begin{equation}
\kappa_6 = \max\Bigl\{2\kappa,\; \frac{X_0^2}{k_1k_5Y_0^2|X_1|}\Bigr\} ,
\label{eq:kappa6}
\end{equation}
is positive definite and radially unbounded in its six arguments, and along the
closed-loop solutions
\begin{equation}
\begin{aligned}
\dot V_6 \le{}& -k_1\rho_a^2\cos^2\gamma_a - \frac{u_b^2}{k_1}
 - \frac{v_b^2}{k_1|X_1|}\\
& - \frac{2k_1k_2\,\delta_a^2}{\sqrt{1+4k_2^2\delta_a^2}}
 - \frac{2k_1k_4}{k_3}\,z^2 - 2k_5\bigl(r_b-\omega^\star\bigr)^2 .
\end{aligned}
\label{eq:V6dot}
\end{equation}
For every initial condition with $\rho_a(0)>0$, the solution exists and is
unique for all $t\ge0$, satisfies $\rho_a(t)>0$, and
\begin{equation}
\begin{aligned}
&\Upsilon_6(t) \le \beta_6\bigl(\Upsilon_6(0),t\bigr), \\
&\Upsilon_6 = \rho_a + |\delta_a| + |\gamma_a| + |u_b| + |v_b| + |r_b| ,
\end{aligned}
\label{eq:KL6}
\end{equation}
for a class $\mathcal{KL}$ function $\beta_6$ that does not depend on the
initial condition. In particular, \eqref{eq:objective} holds and
$(\tau_u,\tau_r)\to0$.
\label{thm:six}
\end{theorem}

For $m_{23}=0$, by \eqref{eq:coeff-m23} and \eqref{eq:fu}, the surge
law \eqref{eq:tau-u} is
\begin{equation}
\tau_u = \frac{d_{22}}{m_{11}}\,k_1\rho_a\cos\gamma_a
 - \frac{d_{22}-d_{11}}{m_{11}}\,u_b .
\label{eq:tau-u-m23}
\end{equation}

\begin{proof}[Proof of Theorem \ref{thm:six}]
By \eqref{eq:udot} and \eqref{eq:tau-u}, $u_b$ obeys the controller equation
of \eqref{eq:fb-ua}, so \eqref{eq:rho-dot}, \eqref{eq:energy} and
\eqref{eq:E-bound} hold without change. Let $\varsigma=r_b-\omega^\star$. By
\eqref{eq:a-delta} and \eqref{eq:a-gamma},
$\dot\delta_a=\tfrac{k_1}{2}\sin(2\gamma_a)$ and
$\dot\gamma_a=\tfrac{k_1}{2}\sin(2\gamma_a)-r_b$. Since
$\dot\varsigma = \dot r_b-\dot\omega^\star
 = f_r+\tau_r
   -\tfrac{\partial\omega^\star}{\partial\delta_a}\dot\delta_a
   -\tfrac{\partial\omega^\star}{\partial\gamma_a}\dot\gamma_a$,
substituting \eqref{eq:tau-r} cancels all terms except $-k_5\varsigma$
and $\tfrac{k_1}{k_3}z$, giving
\begin{equation}
\dot\varsigma = -k_5\,\varsigma + \frac{k_1}{k_3}\,z .
\label{eq:sigma-dot}
\end{equation}
Relative to the closed loop of Theorem \ref{thm:four}, $\dot\gamma_a$ contains
the additional term $-\varsigma$, so
$\dot V_{\delta\gamma}=-W-2(k_1/k_3)z\varsigma$ with $V_{\delta\gamma}$ and $W$
as in \eqref{eq:W}. With \eqref{eq:sigma-dot}, the terms in $z\varsigma$ cancel
in
\begin{equation}
\frac{d}{dt}\bigl(V_{\delta\gamma}+\varsigma^2\bigr) = -W - 2k_5\,\varsigma^2 .
\label{eq:V1dot}
\end{equation}
Since $r_b^2\le2(\omega^\star)^2+2\varsigma^2$, the bound on $|\omega^\star|$ in
the proof of Theorem \ref{thm:four} and \eqref{eq:kappa6} give
\eqref{eq:r-bound} with $\kappa_6$, $V_{\delta\gamma}+\varsigma^2$ and
$W+2k_5\varsigma^2$ in place of $\kappa$, $V_{\delta\gamma}$ and $W$. Adding
\eqref{eq:rho-dot}, \eqref{eq:E-bound} and $\Phi_6'$ times \eqref{eq:V1dot}
gives \eqref{eq:V6dot}. The remaining claims follow as in the proof of Theorem
\ref{thm:four}; the inputs \eqref{eq:tau-u}, \eqref{eq:tau-r} are continuous
functions of the state that vanish at the origin, since $f_u$ and $f_r$ do.
\end{proof}

\begin{theorem}[Six-state problem solved by BoLSA]
Let Assumption \ref{as:coeff} hold, and let $k_1,k_2,k_3,k_5,c>0$ with
$k_1k_3\ge k_2^2$. Consider the six-state model
\eqref{eq:p-rho}--\eqref{eq:rdot} under $\tau_u$ in \eqref{eq:tau-u} and
\begin{equation}
\begin{aligned}
\tau_r ={}& - f_r(u_b,v_b,r_b)
 + \frac{k_1}{2}\sin(2\gamma_a)\,\frac{\partial\omega^\circ}{\partial\delta_a}\\
 & + \Bigl(\frac{k_1}{2}\sin(2\gamma_a) - r_b\Bigr)
   \frac{\partial\omega^\circ}{\partial\gamma_a}\\
 & + \frac{1}{2c}\,\frac{\partial V_{\delta\gamma}}{\partial\gamma_a}
 - k_5\bigl(r_b-\omega^\circ\bigr),
\end{aligned}
\label{eq:tau-r-B}
\end{equation}
with $\omega^\circ$, $V_{\delta\gamma}$ and $q$ as in \eqref{eq:omega-B} and
\eqref{eq:Vdg-B}, $\gamma_a\in(-\pi,\pi)$ and $\delta_a\in\mathbb{R}$. The function
\begin{equation}
V_B = \rho_a^2 + \frac{X_1^2u_b^2+v_b^2}{k_1Y_0X_1}
 + \lambda\Bigl(V_{\delta\gamma} + c\bigl(r_b-\omega^\circ\bigr)^2\Bigr)
\label{eq:VB}
\end{equation}
where
\begin{equation}
\lambda = \max\Bigl\{2\lambda_4-1,\; 1+\frac{2\varepsilon}{ck_5}\Bigr\},
\label{eq:lambda-B}
\end{equation}
with $\lambda_4$ and $\varepsilon$ as in \eqref{eq:lambda4},
satisfies, along the closed-loop solutions,
\begin{equation}
\begin{aligned}
\dot V_B \le{}& -k_1\rho_a^2\cos^2\gamma_a - \frac{u_b^2}{k_1}
 - \frac{v_b^2}{k_1|X_1|}\\
 & - \frac{\lambda}{2}\Bigl[8k_1k_2\tan^2\frac{\gamma_a}{2}
 + \frac{3k_2}{2}\Bigl(\delta_a+2q\tan\frac{\gamma_a}{2}\Bigr)^2\\
 &\qquad + 2ck_5\bigl(r_b-\omega^\circ\bigr)^2\Bigr] .
\end{aligned}
\label{eq:VBdot}
\end{equation}
For every initial condition with $\rho_a(0)>0$ and $|\gamma_a(0)|<\pi$, the
solution exists and is unique for all $t\ge0$, satisfies $\rho_a(t)>0$ and
$|\gamma_a(t)|<\pi$, and \eqref{eq:KL6} holds with $|\gamma_a|$ replaced by
$2|\tan(\gamma_a/2)|$. In particular, \eqref{eq:objective} holds and
$(\tau_u,\tau_r)\to0$.
\label{cor:bolsa}
\end{theorem}

\begin{proof}[Proof of Theorem \ref{cor:bolsa}]
With
$\varsigma=r_b-\omega^\circ$, the closed loop gives $\dot\varsigma=-k_5\varsigma
+\frac{1}{2c}\partial V_{\delta\gamma}/\partial\gamma_a$, and $\dot\gamma_a$
contains the additional term $-\varsigma$ relative to \citet[Eq.~(8)]{part1}, so
by \citet[Eqs.~(50),~(57)]{part1}
\begin{equation}
\begin{aligned}
&\frac{d}{dt}\bigl(V_{\delta\gamma}+c\,\varsigma^2\bigr)
\le -W_B, \\
&W_B = 8k_1k_2\tan^2\frac{\gamma_a}{2}
 + \frac{3k_2}{2}\Bigl(\delta_a+2q\tan\frac{\gamma_a}{2}\Bigr)^2\\
&\qquad + 2ck_5\varsigma^2 .
\end{aligned}
\label{eq:VBdg}
\end{equation}
Since $r_b^2\le2(\omega^\circ)^2+2\varsigma^2$, \eqref{eq:r-bound-4} gives
\begin{equation}
\begin{aligned}
r_b^2 \le{}& \bigl(16(k_1+k_2)^2+32k_1k_3\bigr)s^2\\
 & + 8k_3^2(\delta_a+2qs)^2 + 2\varsigma^2 ,
\qquad s=\tan\frac{\gamma_a}{2}.
\end{aligned}
\label{eq:r-bound-B}
\end{equation}
By \eqref{eq:lambda-B}, the last term of \eqref{eq:E-bound} is then at most
$\frac{\lambda}{2}W_B$. Adding \eqref{eq:rho-dot}, \eqref{eq:E-bound} and
$\lambda$ times \eqref{eq:VBdg} gives \eqref{eq:VBdot}. Since $V_{\delta\gamma}\to\infty$ as $|\gamma_a|\to\pi$ and
$V_B$ is nonincreasing, $|\gamma_a(t)|<\pi$. The remaining claims follow as in
the proof of Theorem \ref{thm:four}.
\end{proof}

{Remark \ref{rem:sense} and Proposition \ref{prop:dir} apply in the
same way to Theorems \ref{cor:bolsa4}, \ref{thm:six} and \ref{cor:bolsa}, with
$|r_b|$ added on both sides of \eqref{eq:KL-O} for the six-state model: the proof
of Proposition \ref{prop:dir} uses only \eqref{eq:fb-ua}, which $u_b$ obeys
under \eqref{eq:tau-u}, and $\gamma_a\to0$, $r_b\to0$.}

%=====================================================================
\begin{figure*}[t]
\centering
\includegraphics[width=0.82\textwidth]{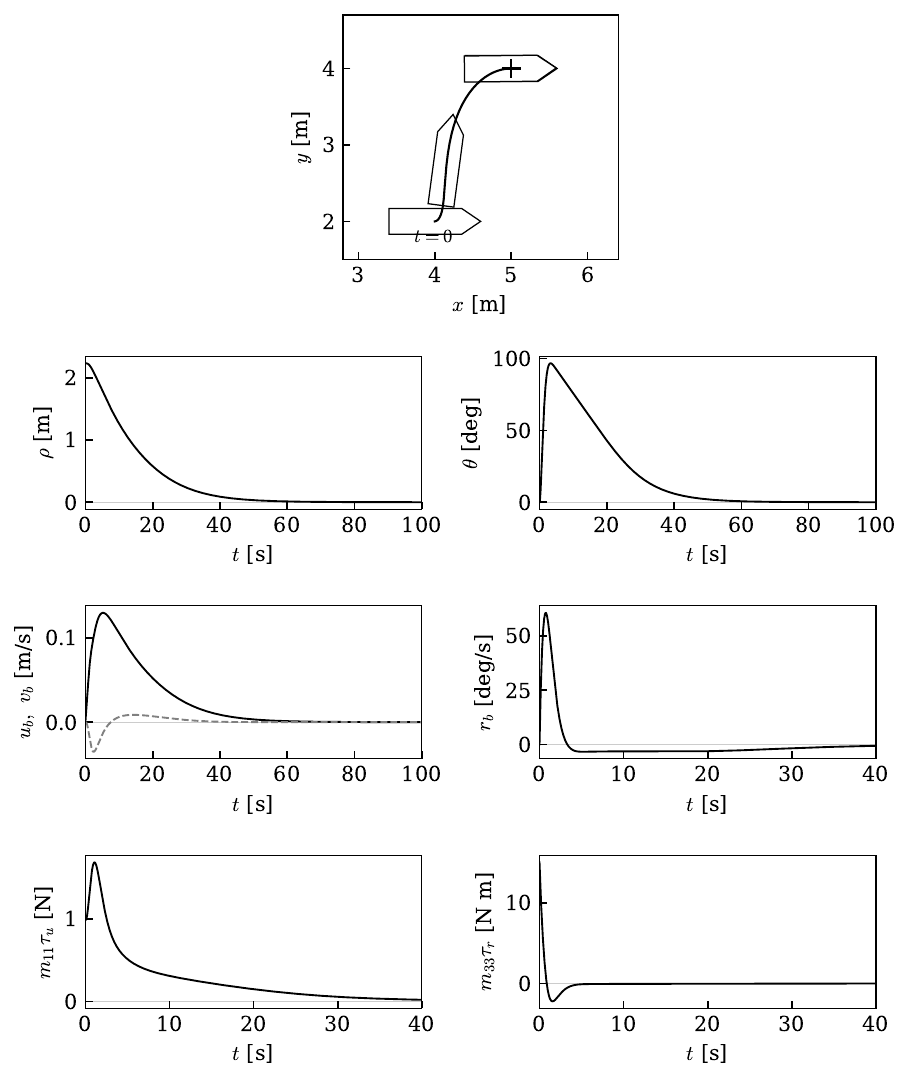}
\caption{Theorem \ref{thm:six} on the maneuver of
\citet[Eqs.~(26),~(30)]{pettersen04}, $d_{23}=d_{32}=0$. Top: path of $O$, with
the hull at $t=0$, $8$ and $60$\,s; the desired pose is shaded. Below: distance
$\rho$, heading error $\theta$, surge speed $u_b$ (solid) and sway speed $v_b$
(dashed), yaw rate $r_b$, surge force $m_{11}\tau_u$ and yaw moment
$m_{33}\tau_r$.}
\label{fig:bench}
\end{figure*}

\begin{figure*}[t]
\centering
\includegraphics[width=0.82\textwidth]{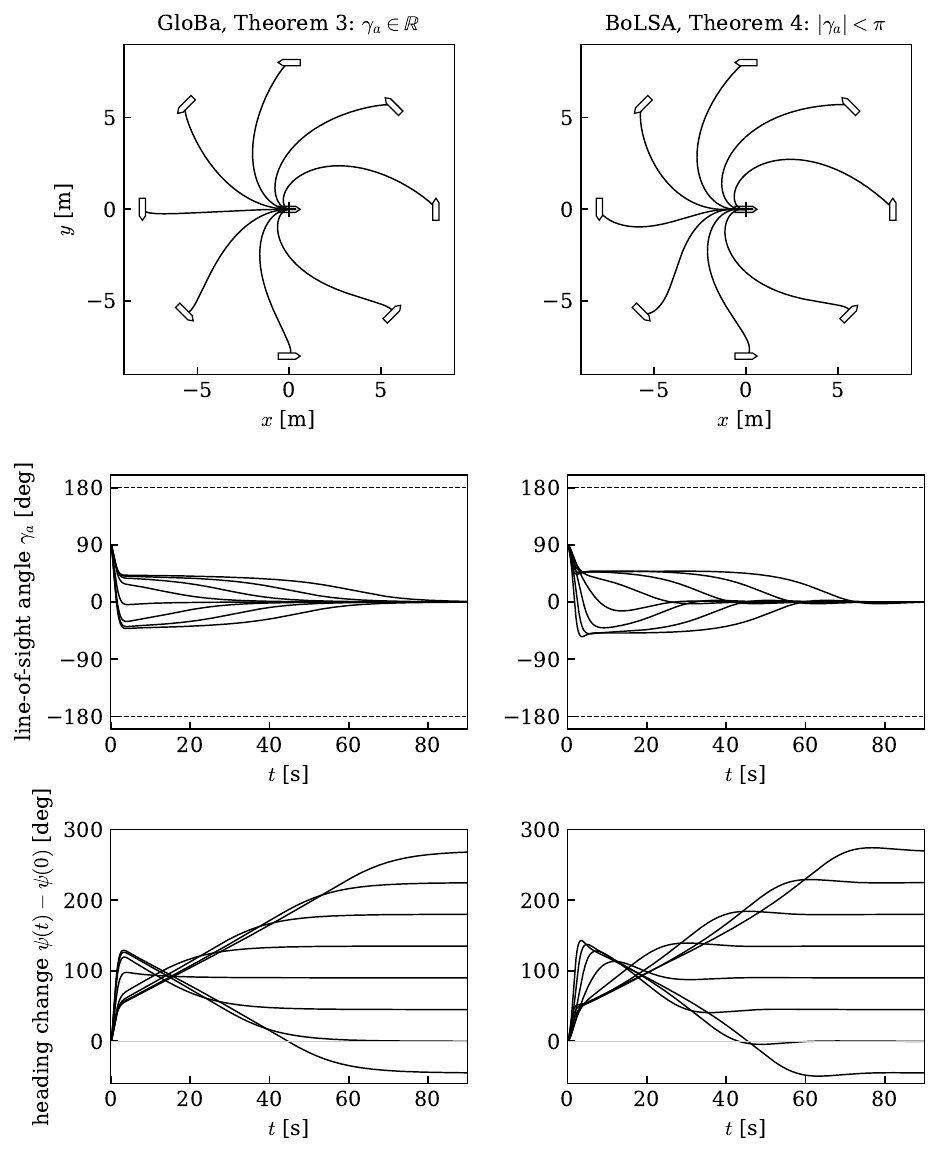}
\caption{Eight maneuvers of the six-state model with inputs $(\tau_u,\tau_r)$,
from rest at distance $8$\,m, $d_{23}=d_{32}=0$. Left: GloBa, Theorem
\ref{thm:six}. Right: BoLSA, Theorem \ref{cor:bolsa}. Initial angles chosen for each start
as described in the text. Top: paths of $O$, with the hull at $t=0$; the
desired pose is shaded. Middle: line-of-sight angle $\gamma_a$; the dashed lines
are $\pm180^\circ$. Bottom: heading change $\psi(t)-\psi(0)$.}
\label{fig:fan}
\end{figure*}

\begin{figure*}[t]
\centering
\begin{minipage}[t]{0.48\textwidth}
\centering
\includegraphics[width=\linewidth]{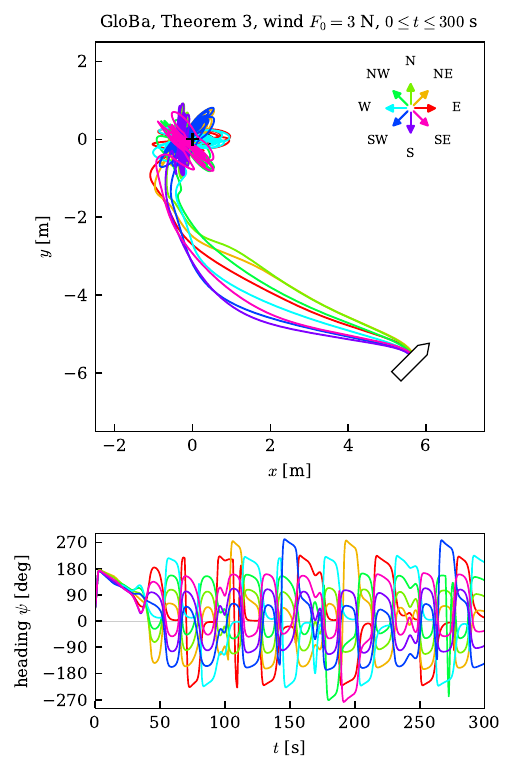}
\caption{Theorem \ref{thm:six} (GloBa) from the start at $(5.66,-5.66)$\,m of
Fig.~\ref{fig:fan}, $d_{23}=d_{32}=0$, under the wind force $F(t)$ in eight
directions, colored by the direction toward which the wind blows. Top: paths of
$O$; the desired pose is shaded. Bottom: heading $\psi$.}
\label{fig:wind}
\end{minipage}\hfill
\begin{minipage}[t]{0.48\textwidth}
\centering
\includegraphics[width=\linewidth]{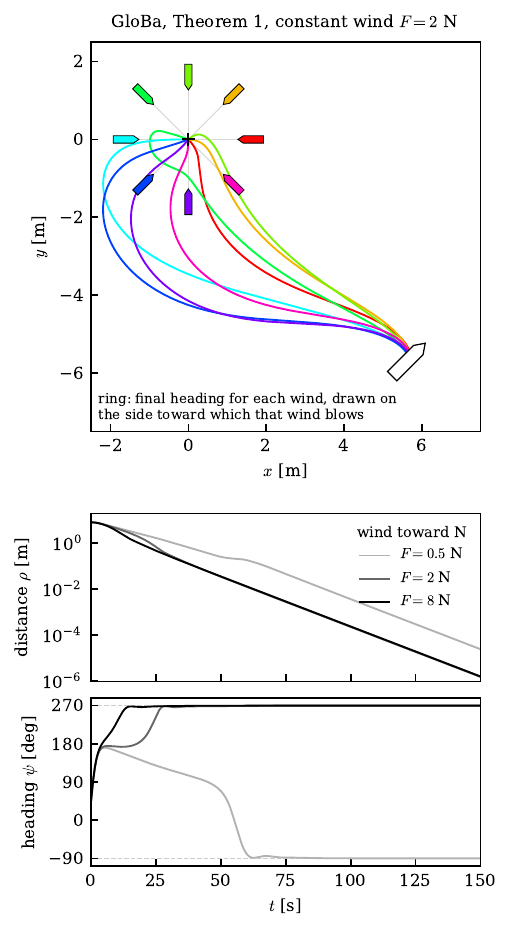}
\caption{Theorem \ref{thm:four} (GloBa, four-state model) from the start at
$(5.66,-5.66)$\,m of Fig.~\ref{fig:fan}, under a constant wind force $F$. Top:
$F=2$\,N in eight directions, colored by the direction toward which the wind
blows; the hulls on the ring show the final heading for each wind. Middle and
bottom: wind toward north with $F=0.5$, $2$ and $8$\,N; distance $\rho$ of $O$
from the desired position and heading $\psi$.}
\label{fig:const}
\end{minipage}
\end{figure*}

\section{Simulations of GloBa and BoLSA on six-state model}\label{sec:sim}

The vessel data are those of Cybership~I in \citet[Eqs.~(33)--(35)]{pettersen04},
\begin{equation}
\begin{aligned}
&m_{11}=19.0, && m_{22}=35.2, && m_{33}=4.2, && m_{23}=0,\\
&d_{11}=4.0, && d_{22}=10.0, && d_{33}=1.0, &&
\end{aligned}
\label{eq:cybership}
\end{equation}
in SI units, so that $X_1/Y_0=1.9$\,s and $-1/Y_0=3.52$\,s by \eqref{eq:coeff-m23}.
The off-diagonal damping is $d_{23}=d_{32}=0$ ($X_0=0$) or $d_{23}=d_{32}=0.3$
($X_0/Y_0=0.03$\,m). The gains are $k_1=0.1$, $k_2=1$, $k_3=0.1$, $k_4=0.5$,
$k_5=2$. The model is \eqref{eq:p-rho}--\eqref{eq:rdot} with
\eqref{eq:fu}--\eqref{eq:fr}, and the feedback is \eqref{eq:tau-u},
\eqref{eq:tau-r}.

Fig.~\ref{fig:bench} uses the desired pose and initial condition of the
experiment in \citet[Eqs.~(26),~(30)]{pettersen04}: desired pose $(5,4)$\,m with
heading $0$, vessel at rest at $(4,2)$\,m with heading $0$. The distance $\rho$
is below $5$\,cm at $t=46$\,s and below $1$\,cm at $t=62$\,s, and the function
$V_6$ decreases monotonically. With $d_{23}=d_{32}=0.3$ the path differs from
the one shown by at most $4$\,cm. The yaw moment is largest at $t=0$, where the
yaw rate $r_b=0$ differs from $\omega^\star$.

Fig.~\ref{fig:fan} compares GloBa (Theorem \ref{thm:six}) with BoLSA (Theorem
\ref{cor:bolsa}) on the six-state model \eqref{eq:p-rho}--\eqref{eq:rdot}, driven
by the surge force and the yaw moment, on eight maneuvers with $d_{23}=d_{32}=0$, each from rest at
distance $8$\,m with heading perpendicular to the line of sight. The gains of
Theorem \ref{cor:bolsa} are $k_1=0.1$, $k_2=0.2$, $k_3=0.4$, $k_5=1$, $c=20$.
In all sixteen runs $\rho<5$\,cm within $85$\,s and the respective Lyapunov
function decreases monotonically. With
$d_{23}=d_{32}=0.3$ and with $d_{23}=d_{32}=2$ ($X_0/Y_0=0.2$\,m), the feedback of
Theorem \ref{cor:bolsa} parks the vessel from the same eight starts within
$85$\,s, with $V_B$ decreasing monotonically ($\lambda=1.57$ and $26.2$).

\emph{Initial angles.} For each start, $\delta_a(0)$ and $\theta(0)$ are
chosen among the values that differ by multiples of $2\pi$ so that the angular
part of the Lyapunov function, $\delta_a^2+(k_1/k_3)z^2$ for GloBa and
$V_{\delta\gamma}$ for BoLSA, is smallest. Every start then has
$\gamma_a(0)=90^\circ$. The start at $(8,0)$\,m lies on the line ahead of the
desired pose, where $\delta_a(0)=180^\circ$ and $\delta_a(0)=-180^\circ$ describe the
same position. For the vessel heading north, the choice gives
$\delta_a(0)=-180^\circ$, and the vessel passes north of $O_d$; the value $180^\circ$
would send it south. A single convention, $\delta_a(0),\theta(0)\in(-\pi,\pi]$ for
all starts, gives $\gamma_a(0)=-270^\circ$ for the GloBa start at
$(5.66,5.66)$\,m, and that vessel turns $310^\circ$ clockwise in the first $4$\,s.

\emph{Initial turn.} In all sixteen maneuvers $|\gamma_a(t)|\le90^\circ$, so no
maneuver winds. Every start is perpendicular to the line of sight, so each vessel
first turns toward it, always counterclockwise. In the first $10$\,s the heading
changes by $67^\circ$--$143^\circ$ under BoLSA and by $74^\circ$--$129^\circ$
under GloBa: about $90^\circ$ to face the line of sight, and up to $53^\circ$
more, to a line-of-sight angle that is held while $\delta_a$ decreases toward
$0$. The desired position is approached from behind, $\delta_a\to0$, and the
extra turn is the start of the curved path that leads there. Over all eight
starts, BoLSA gives yaw rate at most $81^\circ$/s and yaw moment at most
$9.8$\,N\,m, and GloBa gives at most $82^\circ$/s and $19.3$\,N\,m.

\begin{figure}[t]
\centering
\includegraphics[width=\columnwidth]{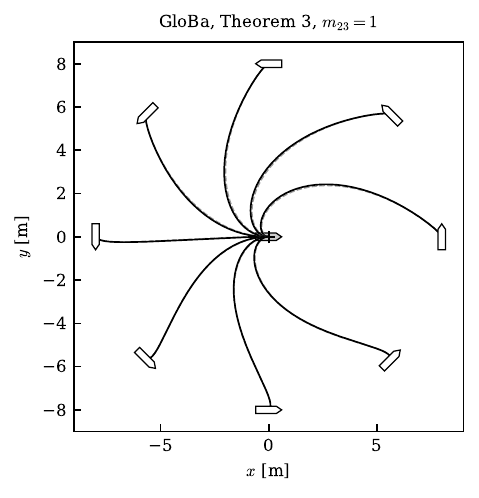}
\caption{Theorem \ref{thm:six} (GloBa) from the eight starts of
Fig.~\ref{fig:fan} on the hull with $m_{23}=1$, which violates Assumption
\ref{as:coeff}. Solid: paths of $O$; dashed grey: paths with $m_{23}=0$ from
Fig.~\ref{fig:fan}. The desired pose is shaded.}
\label{fig:m23}
\end{figure}

\emph{Wind.} Fig.~\ref{fig:wind} repeats the GloBa maneuver from the start at
$(5.66,-5.66)$\,m of Fig.~\ref{fig:fan} under the zero-mean almost periodic
wind force
$F(t)=3\bigl(\sin\omega t+0.7\sin(\sqrt2\,\omega t+1.3)+0.5\sin(\sqrt5\,\omega t+2.9)\bigr)/2.2$\,N,
$\omega=0.21$\,rad/s, acting at $O$ in each of eight fixed directions. This serves as a stress test of the controller's robustness to oscillatory disturbances.
One would expect a crosswind to be the hard case for a vessel that cannot push
sideways, and a head or tail wind to be the easy one. Instead, the vessel holds
its position almost equally well in every wind, because it gives up its heading
and keeps swinging its hull around, so every wind ends up meeting it from all
sides.

\emph{Constant wind.} Fig.~\ref{fig:const} shows the four-state model under the
feedback of Theorem \ref{thm:four}, from the same start, with a constant force
$F$ in a fixed direction $\varphi$. The surge speed is commanded in this
model, so the wind acts only through the sway term of \eqref{eq:vdot-F},
which for $m_{23}=0$ is $(F/m_{22})\sin(\varphi-\theta)$. In every direction and at every strength, the
vessel parks exactly at the desired position, with its bow pointing into the
wind. The heading error is therefore the angle between the desired heading and
the direction from which the wind comes, and it does not shrink as the wind
weakens: a light wind leaves the vessel turned by $90^\circ$ just as a strong
one does. This rules out input-to-state stability with respect to the wind,
which would require every error, the heading error included, to become small
when the wind is small. For a vessel holding its position, however, this is the
expected and even desirable outcome: with the bow into the wind, the sideways
push that the vessel cannot counteract disappears, which is how a seaman keeps
station in steady weather.

\emph{Hull coupling.} Assumption \ref{as:coeff} holds for real hulls only
approximately: $m_{22}>m_{11}$ for all hulls whose length exceeds their beam, and the sway--yaw coupling $m_{23}$ is
nonzero for a hull that is not symmetric fore and aft. Fig.~\ref{fig:m23}
repeats the GloBa maneuvers of Fig.~\ref{fig:fan} with $m_{23}=1$ in
\eqref{eq:cybership}. The sway equation then contains the term $Y_1u_bv_b$ of
\eqref{eq:Yu}, with $Y_1=0.11$, which the feedback \eqref{eq:tau-u},
\eqref{eq:tau-r} ignores; its coefficients $X_0$, $X_1$, $Y_0$ are computed by
\eqref{eq:Xu}, \eqref{eq:Yu} with $m_{23}=1$. All eight maneuvers reach
$\rho<5$\,cm within $76$\,s, and their paths differ from those with $m_{23}=0$
by at most $8$\,cm.

%=====================================================================
\section{Scope and prior results}

\begin{remark}[Initial conditions]
The set $\{\rho_a=0\}$ excluded in Theorems \ref{thm:four} and \ref{thm:six} is
$\{a=a_d\}$, which has codimension two.
\end{remark}

\begin{remark}[Other steering laws of {\citet{part1}}]\label{rem:other}
If $X_0=0$ and \eqref{eq:fb-r} is replaced by \citet[Eq.~(7)]{part1} with any of
the eight steering laws of \citet[Thms.~1--8]{part1}, with angular CLF
$V_{\delta\gamma}$, then on the corresponding state space
\begin{equation}
\begin{aligned}
&\frac{d}{dt}\Bigl(\rho_a^2 + \frac{X_1^2u_b^2+v_b^2}{k_1Y_0X_1}
 + V_{\delta\gamma}\Bigr)\\
&\le -k_1\rho_a^2\cos^2\gamma_a - \frac{u_b^2}{k_1}
 - \frac{v_b^2}{k_1|X_1|} + \dot V_{\delta\gamma} ,
\end{aligned}
\label{eq:eight}
\end{equation}
with $\dot V_{\delta\gamma}$ bounded by the negative definite function given in
the proof of the respective theorem of \citet{part1}. This follows from
\eqref{eq:rho-dot} and \eqref{eq:E-bound}. The designs of \citet{part2} have not
been examined for \eqref{eq:a-rho}--\eqref{eq:a-v}.
\end{remark}

\begin{remark}[Prior result]\label{rem:prior}
For the six-state model with $m_{23}=d_{23}=d_{32}=0$, which satisfies Assumption
\ref{as:coeff} with $X_0=0$, the objective is attained in \citet{mazenc} by a
feedback that is smooth, time-varying and periodic in $t$, built by
backstepping, with global uniform asymptotic stability;
experiments are in \citet{pettersen04}. That construction is not Pomet's, which
applies to driftless systems. For the same coefficients, the displacement of
$a$ from $O$ has the components $(m_{11}/d_{22})u_b$ and $(m_{22}/d_{22})v_b$,
and the components of the position of $a$ along the body axes are the states
$-\mu$ and $Z_2$ of \citet[Eqs.~(7),~(9)]{mazenc} and, up to sign, two of the
transformed states of \citet{dongguo}. In those works they are coordinates for a
time-varying design; here they are the position of a point whose velocity points
along the heading, used with autonomous feedback in polar coordinates.
\end{remark}

\begin{remark}[Coefficient of $u_bv_b$]\label{rem:Y1}
If the sway acceleration contains the additional term $Y_1u_bv_b$, as in
\eqref{eq:Yu} with $m_{23}(m_{22}-m_{11})\neq0$, the point $a$ has
the lateral speed $-(Y_1/Y_0)u_bv_b$, and no point displaced from $O$ by
continuous functions of $(u_b,v_b,r_b)$ has zero lateral speed for every input.
This case is not covered.
\end{remark}

%=====================================================================
\section{Conclusions}

The underactuated surface vessel with sideways slip is stabilized in polar
coordinates by time-invariant feedback, first with the surge speed and yaw rate
as inputs (Theorems \ref{thm:four} and \ref{cor:bolsa4}) and then with the surge
force and yaw moment (Theorems \ref{thm:six} and \ref{cor:bolsa}). The
zero-sideslip point makes the steering laws of \citet{part1} applicable to the
vessel. The CLF construction extends to hulls with fore-aft asymmetric sway--yaw damping, excluded in~\citet{mazenc,dongguo} by the assumption of diagonal damping. Unlike the periodic time-varying designs of~\citet{mazenc,dongguo}, the time-invariant controller yields smooth, non-oscillatory trajectories---an important advantage for docking.

For the same model of an underactuated vessel (apart from the diagonal inertia
and the fully coupled damping matrix of \citet{wang23}, in place of the inertia
with the coupling $m_{23}$ and the sway--yaw damping block used here, and apart
from the scaling of the inputs), whereas \citet{wang23} steers the vessel toward
the unknown minimizer of a field using only measurements of the field's value,
and guarantees, with time-varying dither feedback on the surge force and a
constant yaw torque that keeps the vessel turning, only semi-global practical
convergence of the position, the present paper stabilizes the vessel at a known
pose using the full state, and guarantees, with time-invariant feedback through
both the surge force and the yaw moment, global asymptotic convergence of the
position, heading, and velocities to rest. The feedback laws of
\citet{mazenc} and \citet{wang23} share one feature, a periodic dependence on time
through which the vessel advances by repeated forward and backward surge
motions, as observed in the experiments of \citet{pettersen04} and \citet{wang23};
the feedback laws of this paper are time-invariant and impose no such periodic
reversals.

Of the eight steering laws of \citet{part1}, two more have properties relevant to
a vessel: BAR-FLi \citep[Thm.~8]{part1} keeps $|\delta_a|<\pi$, and BAgAl
\citep[Thm.~4]{part1} keeps $|\delta_a|<\pi$ and $|\gamma_a|<\pi$, so that the
vessel never crosses the line ahead of the desired pose and, under BAgAl, the
heading error satisfies $|\theta(t)|<2\pi$. For $X_0=0$ both are covered by
Remark \ref{rem:other}. For $X_0\neq0$, the yaw rates of these laws grow like
$\tan^3(\delta_a/2)$ as $\delta_a\to\pm\pi$, and the angular dissipation grows
more slowly, so the argument of Theorem \ref{thm:four} requires a weight
$\Phi$ on the angular CLF with
$\Phi'(V_{\delta\gamma})$ growing like $V_{\delta\gamma}^{3/2}$ for BAR-FLi and
like $V_{\delta\gamma}^{2/3}$ for BAgAl, in place of $V_{\delta\gamma}^{1/2}$ for
GloBa and of a constant for BoLSA.

Relative to GloBa, BAR-FLi keeps the assignment of the local eigenvalues
\citep[Prop.~6]{part1} and adds the bound on $\delta_a$, at the cost of the
starts with $\delta_a(0)=\pm\pi$, which it excludes, and of yaw-rate commands
that are unbounded near the line ahead of the desired pose. Relative to BoLSA,
BAgAl adds the bound on $\delta_a$ and hence on $\theta$, at the same costs; its
first maneuver from a start near that line is a fast turn while the forward
speed is not small \citep[Fig.~6]{part1}, which for the vessel requires a large
yaw moment. Both designs require longer proofs than Theorems
\ref{thm:four}--\ref{cor:bolsa}, and they are not pursued here because of the
length of the paper.

Simulations of GloBa and BoLSA on the Cybership~I model confirm the
theoretical results: both laws park the six-state vessel at a prescribed pose
from eight starting positions covering all approach angles, with the respective
Lyapunov functions decreasing monotonically throughout.
Under a zero-mean almost periodic wind force acting in eight fixed inertial
directions, GloBa holds the desired position equally well regardless of wind
direction, exploiting the vessel's freedom to adjust its heading.
Simulations with a constant wind force~$F$ show that the vessel parks at the
desired position with its bow pointing into the wind.

\appendix

\section{Proof of Proposition \ref{prop:a}}\label{app:proof}

Let $F=\rho_a\cos\gamma_a$ and $G=\rho_a\sin\gamma_a$ denote the left sides of
\eqref{eq:F} and \eqref{eq:G}, and let $\theta=\delta-\gamma$. By
\eqref{eq:p-delta} and \eqref{eq:p-gamma}, $\dot\theta=r_b$. By
\eqref{eq:p-rho} and \eqref{eq:p-gamma},
\begin{subequations}
\begin{align}
\frac{d}{dt}\bigl(\rho\cos\gamma\bigr) &= -u_b + r_b\,\rho\sin\gamma,
 \label{eq:F0}\\
\frac{d}{dt}\bigl(\rho\sin\gamma\bigr) &= -v_b - r_b\,\rho\cos\gamma .
 \label{eq:G0}
\end{align}
\end{subequations}
Differentiating \eqref{eq:G} with \eqref{eq:G0}, \eqref{eq:p-v} and
$\dot\theta=r_b$, the terms in $v_b$ cancel and
\begin{equation}
\dot G = -r_b\Bigl(\rho\cos\gamma - \frac{X_0+X_1u_b}{Y_0}
 + \frac{X_0}{Y_0}\cos\theta\Bigr) = -r_b\,F .
\label{eq:Gdot}
\end{equation}
Differentiating \eqref{eq:F} with \eqref{eq:F0} and $\dot\theta=r_b$, and
substituting $\rho\sin\gamma$ from \eqref{eq:G},
\begin{equation}
\dot F = -\Bigl(u_b + \frac{v_br_b}{Y_0} + \frac{X_1}{Y_0}\dot u_b\Bigr) + r_b\,G
 = -u_a + r_b\,G .
\label{eq:Fdot}
\end{equation}
From \eqref{eq:Gdot} and \eqref{eq:Fdot},
$\rho_a\dot\rho_a=F\dot F+G\dot G=-u_aF$, which is \eqref{eq:a-rho}, and
$\rho_a^2\dot\gamma_a=F\dot G-G\dot F=u_aG-r_b\rho_a^2$, which is
\eqref{eq:a-gamma}. Equation \eqref{eq:a-delta} follows from
\eqref{eq:delta-a}, \eqref{eq:a-gamma} and $\dot\theta=r_b$. The relations
\eqref{eq:F}, \eqref{eq:G} contain $(\rho,\delta,\gamma)$ only through
$\rho\cos\gamma$, $\rho\sin\gamma$ and $\delta-\gamma$, which remain defined
where $\rho=0$. \hfill$\Box$

\section{Cartesian model, hull coefficients and polar variables}\label{app:cart}

The polar model \eqref{eq:p-rho}--\eqref{eq:rdot} is the Cartesian model of
\citet[Eq.~(1)]{haraldsen21} written in polar variables. With $(x_b^n,y_b^n)$
the position of $O$ and $\psi_b^n$ the heading,
\begin{subequations}
\begin{align}
\dot x_b^n &= u_b\cos\psi_b^n - v_b\sin\psi_b^n, \notag\\
\dot y_b^n &= u_b\sin\psi_b^n + v_b\cos\psi_b^n, \qquad
\dot\psi_b^n = r_b, \label{eq:cart-kin}\\
\dot u_b &= f_u(u_b,v_b,r_b)+\tau_u, \qquad
\dot r_b = f_r(u_b,v_b,r_b)+\tau_r, \notag\\
\dot v_b &= X(u_b)\,r_b + Y(u_b)\,v_b . \label{eq:cart-dyn}
\end{align}
\end{subequations}
The body axes point forward and to starboard. The origin $O$ is the pivot point
\citep{fossen14}: the point on the centerline at which the actuators produce no
sway acceleration, so that no input enters the sway equation. The sideslip
angle $\atantwo(v_b,u_b)$ of $O$ is nonzero whenever $v_b\ne0$; the
zero-sideslip point $a$ of Definition \ref{def:a} is a different point, whose
velocity has zero sideslip angle at every instant.

The hull parameters enter through the mass matrix (rigid-body mass, inertia and added
mass) and the linear damping matrix \citep{fossen11},
\begin{equation}
M=\begin{bmatrix} m_{11} & 0 & 0\\ 0 & m_{22} & m_{23}\\ 0 & m_{23} & m_{33}\end{bmatrix},
\qquad
D=\begin{bmatrix} d_{11} & 0 & 0\\ 0 & d_{22} & d_{23}\\ 0 & d_{32} & d_{33}\end{bmatrix}.
\label{eq:MD}
\end{equation}
With $\Delta=m_{22}m_{33}-m_{23}^2>0$, the functions in \eqref{eq:cart-dyn}
are \citep{haraldsen26}
\begin{subequations}
\begin{align}
f_u &= \frac{(m_{22}v_b+m_{23}r_b)\,r_b-d_{11}u_b}{m_{11}}, \label{eq:fu}\\
X(u_b) &= X_0+X_1 u_b, \notag\\
 X_0 &= \frac{d_{33}m_{23}-d_{23}m_{33}}{\Delta}, \notag\\
 X_1 &= \frac{m_{23}^2-m_{11}m_{33}}{\Delta}, \label{eq:Xu}\\
Y(u_b) &= Y_0+Y_1 u_b, \notag\\
 Y_0 &= \frac{-d_{22}m_{33}+d_{32}m_{23}}{\Delta}, \notag\\
 Y_1 &= \frac{(m_{22}-m_{11})\,m_{23}}{\Delta}, \label{eq:Yu}\\
f_r &= \frac{m_{23}d_{22}-m_{22}d_{32}}{\Delta}\,v_b
 + \frac{m_{23}d_{23}-m_{22}d_{33}}{\Delta}\,r_b \notag\\
 &\quad - \frac{(m_{22}-m_{11})\,u_b}{\Delta}\,\bigl(m_{22}v_b+m_{23}r_b\bigr) . \label{eq:fr}
\end{align}
\end{subequations}
The inputs are $\tau_u=b_{11}T_u/m_{11}$ and
$\tau_r=(m_{22}b_{32}-m_{23}b_{22})T_r/\Delta$, where $T_u$ and $T_r$ are the
surge and steering commands and $b_{ij}$ the entries of the actuator map.
The coefficients are constant, and Assumption \ref{as:coeff} is a condition on
\eqref{eq:Xu}, \eqref{eq:Yu}.

With $(x_d,y_d,\psi_d)$ the desired pose, the polar variables are defined as
in \citet[Sec.~II]{part1}:
\begin{subequations}
\begin{align}
\begin{bmatrix} e_x \\ e_y \end{bmatrix}
&=
\begin{bmatrix} \cos\psi_d & \sin\psi_d \\ -\sin\psi_d & \cos\psi_d \end{bmatrix}
\begin{bmatrix} x_b^n - x_d \\ y_b^n - y_d \end{bmatrix}, \notag\\
\theta &= \psi_b^n - \psi_d, \label{eq:cart-error}\\
\rho &= \sqrt{e_x^2 + e_y^2}, \quad
\delta = \atantwo(e_y,e_x) + \pi, \notag\\
\gamma &= \delta - \theta . \label{eq:cart-polar}
\end{align}
\end{subequations}
For $\rho>0$, differentiating \eqref{eq:cart-error}, \eqref{eq:cart-polar} along \eqref{eq:cart-kin}
gives \eqref{eq:p-rho}--\eqref{eq:p-gamma}, and \eqref{eq:p-v}--\eqref{eq:rdot}
are \eqref{eq:cart-dyn} with $Y(u_b)=Y_0$.
With $\hat x_b$ and $\hat y_b$ the forward and starboard unit vectors of the
vessel, $p=[x_b^n,y_b^n]^\top$, $p_d=[x_d,y_d]^\top$ and
$\hat x_d=[\cos\psi_d,\sin\psi_d]^\top$,
\begin{equation}
\begin{aligned}
a &= p + \frac{X(u_b)}{Y_0}\,\hat x_b - \frac{v_b}{Y_0}\,\hat y_b , \qquad
\dot a = u_a\,\hat x_b ,\\
a_d &= p_d + \frac{X_0}{Y_0}\,\hat x_d .
\end{aligned}
\label{eq:cart-a}
\end{equation}
The quantities $\rho_a\cos\gamma_a$ and $\rho_a\sin\gamma_a$ in \eqref{eq:F},
\eqref{eq:G} are the components of $a_d-a$ along $\hat x_b$ and $\hat y_b$.
The angles are real numbers: their values at the initial time are fixed modulo
$2\pi$ by \eqref{eq:cart-polar} and \eqref{eq:F}, \eqref{eq:G}, any
determination is admissible, subject to $|\gamma_a(0)|<\pi$ for BoLSA, and they
are continued continuously in $t$. The
objective \eqref{eq:objective} is the Cartesian objective
$(x_b^n-x_d,\,y_b^n-y_d,\,\psi_b^n-\psi_d,\,v_b,\,u_b,\,r_b)\to0$, with the
heading error taken modulo $2\pi$, because $\rho\to0$ is the convergence of
the position and $\theta\to0$ the convergence of the heading.

\begin{ack}

% The authors' problems, ideas, and results were developed with the assistance of
% Claude in final theorem formulation, proofs, and drafting throughout the paper,
% under the authors' correction and complete verification.
The control problem, the building of the results upon the authors' prior art, the vessel models, and the overall research direction were determined by the authors. Some mathematical derivations were carried out by Claude (Anthropic) under the authors’ continuous direction, correction and careful verification of all results. {The authors thank Kwang Hak Kim and Velimir Todorovski for helpful feedback on the manuscript.}
\end{ack}


\begin{thebibliography}{99}

{\bibitem[Aicardi et~al.(1995)]{aicardi95}
Aicardi, M., Casalino, G., Bicchi, A., \& Balestrino, A. (1995).
Closed loop steering of unicycle-like vehicles via Lyapunov techniques.
\emph{IEEE Robotics \& Automation Magazine}, \emph{2}(1), 27--35.}

\bibitem[Brockett(1983)]{brockett}
Brockett, R.~W. (1983).
Asymptotic stability and feedback stabilization.
In R.~W. Brockett, R.~S. Millman, \& H.~J. Sussmann (Eds.), \emph{Differential geometric control theory} (pp.~181--191). Boston: Birkh\"auser.

\bibitem[Caharija et~al.(2016)]{caharija16}
Caharija, W., Pettersen, K.~Y., Bibuli, M., Calado, P., Zereik, E., Braga, J.,
  Gravdahl, J.~T., S{\o}rensen, A.~J., Milovanović, M., \& Bruzzone, G. (2016).
Integral line-of-sight guidance and control of underactuated marine vehicles:
  Theory, simulations, and experiments.
\emph{IEEE Transactions on Control Systems Technology}, \emph{24}(5), 1623--1642.

\bibitem[Do et~al.(2002)]{do02}
Do, K.~D., Jiang, Z.~P., \& Pan, J. (2002).
Universal controllers for stabilization and tracking of underactuated ships.
\emph{Systems \& Control Letters}, \emph{47}(4), 299--317.

\bibitem[Do et~al.(2003)]{do03}
Do, K.~D., Jiang, Z.~P., \& Pan, J. (2003).
Robust global stabilization of underactuated ships on a linear course:
  State and output feedback.
\emph{International Journal of Control}, \emph{76}(1), 1--17.

\bibitem[Dong and Guo(2005)]{dongguo}
Dong, W., \& Guo, Y. (2005).
Global time-varying stabilization of underactuated surface vessel.
\emph{IEEE Transactions on Automatic Control}, \emph{50}(6), 859--864.

\bibitem[Fossen(2011)]{fossen11}
Fossen, T.~I. (2011).
\emph{Handbook of marine craft hydrodynamics and motion control}.
Chichester: Wiley.

\bibitem[Fossen and Pettersen(2014)]{fossen14}
Fossen, T.~I., \& Pettersen, K.~Y. (2014).
On uniform semiglobal exponential stability (USGES) of proportional line-of-sight guidance laws.
\emph{Automatica}, \emph{50}(11), 2912--2917.

\bibitem[Haraldsen et~al.(2021)]{haraldsen21}
Haraldsen, A., Wiig, M.~S., \& Pettersen, K.~Y. (2021).
Reactive collision avoidance for underactuated surface vehicles using the collision cone concept.
In \emph{Proceedings of the IEEE Conference on Control Technology and Applications} (pp.~619--626).

\bibitem[Haraldsen et~al.(2026)]{haraldsen26}
Haraldsen, A., Wiig, M.~S., \& Pettersen, K.~Y. (2026).
Reactive multiobstacle avoidance for underactuated marine vehicles.
\emph{IEEE Transactions on Control Systems Technology}, \emph{34}(2), 813--828.

\bibitem[Jiang(2002)]{jiang02}
Jiang, Z.~P. (2002).
Global tracking control of underactuated ships by {L}yapunov's direct method.
\emph{Automatica}, \emph{38}(2), 301--309.

\bibitem[Kim et~al.(2025)]{part2}
Kim, K.~H., Todorovski, V., \& Krsti\'c, M. (2025).
Nonholonomic robot parking by feedback---Part II: Nonmodular, inverse optimal, adaptive, prescribed/fixed-time and safe designs.
arXiv:2511.15219.

\bibitem[Kim et~al.(2026)]{threeD}
Kim, K.~H., Todorovski, V., \& Krsti\'c, M. (2026).
Global exponential stabilization of a 3D nonholonomic vehicle in spherical coordinates.
arXiv:2609.24192.

\bibitem[Krsti\'c and Pettersen(2026)]{partII}
Krsti\'c, M., \& Pettersen, K.~Y. (2026).
Control of underactuated surface vessels---Part II: Underway docking.
Submitted to \emph{Automatica}.

\bibitem[Luo and Qi(2021)]{luo21}
Luo, W., \& Qi, X. (2021).
{L2}-gain-based practical stabilization of an underactuated surface vessel.
\emph{Journal of Marine Science and Engineering}, \emph{9}(3), 341.

\bibitem[Ma(2009)]{ma09}
Ma, B. (2009).
Global $\kappa$-exponential asymptotic stabilization of underactuated surface
  vessels.
\emph{Systems \& Control Letters}, \emph{58}(3), 194--201.

\bibitem[Mazenc et~al.(2002)]{mazenc}
Mazenc, F., Pettersen, K.~Y., \& Nijmeijer, H. (2002).
Global uniform asymptotic stabilization of an underactuated surface vessel.
\emph{IEEE Transactions on Automatic Control}, \emph{47}(10), 1759--1762.

\bibitem[Pan et~al.(2026)]{pan26}
Pan, Q., Peng, X., Lv, Y., \& Tian, Y. (2026).
Global exponential stabilization of underactuated surface vessel with
  high-order fully actuated control approach.
\emph{Asian Journal of Control}, \emph{28}(4), 1924--1933.

\bibitem[Pettersen et~al.(2004)]{pettersen04}
Pettersen, K.~Y., Mazenc, F., \& Nijmeijer, H. (2004).
Global uniform asymptotic stabilization of an underactuated surface vessel: Experimental results.
\emph{IEEE Transactions on Control Systems Technology}, \emph{12}(6), 891--903.

\bibitem[Reyhanoglu et~al.(1999)]{reyhanoglu99}
Reyhanoglu, M., van der Schaft, A., McClamroch, N.~H., \& Kolmanovsky, I.
  (1999).
Dynamics and control of a class of underactuated mechanical systems.
\emph{IEEE Transactions on Automatic Control}, \emph{44}(9), 1663--1671.

\bibitem[Todorovski et~al.(2025)]{part1}
Todorovski, V., Kim, K.~H., Astolfi, A., \& Krsti\'c, M. (2025).
Nonholonomic robot parking by feedback---Part I: Modular strict CLF designs.
arXiv:2511.15119.

\bibitem[Todorovski et~al.(2027)]{tractortrailer}
Todorovski, V., Wang, Y., Ren, Z., Leung, J., \& Krsti\'c, M. (2027).
Global exponential stabilization of the unicycle tractor-trailer in polar coordinates.
Submitted to the American Control Conference.

\bibitem[Wang et~al.(2023)]{wang23}
Wang, B., Nersesov, S., Ashrafiuon, H., Naseradinmousavi, P., \& Krsti\'c, M. (2023).
Underactuated source seeking by surge force tuning: Theory and boat experiments.
\emph{IEEE Transactions on Control Systems Technology}, \emph{31}(4), 1649--1662.

\bibitem[Wichlund et~al.(1995)]{pettersen95}
Wichlund, K.~Y., S{\o}rdalen, O.~J., \& Egeland, O. (1995).
Control properties of underactuated vehicles.
In \emph{Proceedings of the IEEE International Conference on Robotics and
  Automation (ICRA)} (pp.~2009--2014). Nagoya: IEEE.

\bibitem[Zhang et~al.(2022)]{zhang22}
Zhang, Z., Tian, L., \& Wu, Y. (2022).
Stabilization of asymmetric underactuated ships with input saturation:
  From underactuated to nonholonomic configuration.
\emph{Ocean Engineering}, \emph{262}, 112177.
\end{thebibliography}
\end{document}